\documentclass[12pt,prd,tightenlines,nofootinbib]{revtex4}
\newcommand{\be}{\begin{equation}}
\newcommand{\ee}{\end{equation}}
\usepackage{bm}
\usepackage{graphics}
\usepackage{rotating}
\usepackage{epsfig}
\begin{document}
\title{\begin{flushright}{\rm\normalsize HU-EP-05/28}\end{flushright}
Relativistic effects in the processes of heavy quark fragmentation}
\author{A. P. 
Martynenko\footnote{E-mail:~apm@physik.hu-berlin.de;~mart@ssu.samara.ru}}
\affiliation{Institut f\"ur Physik, Humboldt--Universit\"at zu Berlin,
Newtonstr. 15, D-12489  Berlin, Germany}
\affiliation{Samara State University, Pavlov Street 1, Samara 443011,
Russia}

\begin{abstract}
In the framework based on the quasipotential method and relativistic quark
model a new covariant expression
for the heavy quark fragmentation amplitude to fragment into the
pseudoscalar and vector $S$-wave heavy mesons is obtained. It contains all
possible relativistic corrections including the terms connected with
the transformation law of the bound state wave function
to the reference frame of the moving meson.
Relativistic corrections of order ${\bf p}^2/m^2$ to the heavy quark
fragmentation distributions into $(\bar c c)$, $(\bar b c)$ and $(\bar b b)$
states are calculated as functions of the longitudinal momentum fraction $z$
and the transverse momentum $p_T$ relative to the jet axis.
\end{abstract}


\maketitle

\immediate\write16{<<WARNING: LINEDRAW macros work with emTeX-dvivers
                    and other drivers supporting emTeX \special's
                    (dviscr, dvihplj, dvidot, dvips, dviwin, etc.) >>}

\newdimen\Lengthunit       \Lengthunit  = 1.5cm
\newcount\Nhalfperiods     \Nhalfperiods= 9
\newcount\magnitude        \magnitude = 1000

\catcode`\*=11
\newdimen\L*   \newdimen\d*   \newdimen\d**
\newdimen\dm*  \newdimen\dd*  \newdimen\dt*
\newdimen\a*   \newdimen\b*   \newdimen\c*
\newdimen\a**  \newdimen\b**
\newdimen\xL*  \newdimen\yL*
\newdimen\rx*  \newdimen\ry*
\newdimen\tmp* \newdimen\linwid*

\newcount\k*   \newcount\l*   \newcount\m*
\newcount\k**  \newcount\l**  \newcount\m**
\newcount\n*   \newcount\dn*  \newcount\r*
\newcount\N*   \newcount\*one \newcount\*two  \*one=1 \*two=2
\newcount\*ths \*ths=1000
\newcount\angle*  \newcount\q*  \newcount\q**
\newcount\angle** \angle**=0
\newcount\sc*     \sc*=0

\newtoks\cos*  \cos*={1}
\newtoks\sin*  \sin*={0}

\catcode`\[=13

\def\rotate(#1){\advance\angle**#1\angle*=\angle**
\q**=\angle*\ifnum\q**<0\q**=-\q**\fi
\ifnum\q**>360\q*=\angle*\divide\q*360\multiply\q*360\advance\angle*-\q*\fi
\ifnum\angle*<0\advance\angle*360\fi\q**=\angle*\divide\q**90\q**=\q**
\def\sgcos*{+}\def\sgsin*{+}\relax
\ifcase\q**\or
 \def\sgcos*{-}\def\sgsin*{+}\or
 \def\sgcos*{-}\def\sgsin*{-}\or
 \def\sgcos*{+}\def\sgsin*{-}\else\fi
\q*=\q**
\multiply\q*90\advance\angle*-\q*
\ifnum\angle*>45\sc*=1\angle*=-\angle*\advance\angle*90\else\sc*=0\fi
\def[##1,##2]{\ifnum\sc*=0\relax
\edef\cs*{\sgcos*.##1}\edef\sn*{\sgsin*.##2}\ifcase\q**\or
 \edef\cs*{\sgcos*.##2}\edef\sn*{\sgsin*.##1}\or
 \edef\cs*{\sgcos*.##1}\edef\sn*{\sgsin*.##2}\or
 \edef\cs*{\sgcos*.##2}\edef\sn*{\sgsin*.##1}\else\fi\else
\edef\cs*{\sgcos*.##2}\edef\sn*{\sgsin*.##1}\ifcase\q**\or
 \edef\cs*{\sgcos*.##1}\edef\sn*{\sgsin*.##2}\or
 \edef\cs*{\sgcos*.##2}\edef\sn*{\sgsin*.##1}\or
 \edef\cs*{\sgcos*.##1}\edef\sn*{\sgsin*.##2}\else\fi\fi
\cos*={\cs*}\sin*={\sn*}\global\edef\gcos*{\cs*}\global\edef\gsin*{\sn*}}\relax
\ifcase\angle*[9999,0]\or
[999,017]\or[999,034]\or[998,052]\or[997,069]\or[996,087]\or
[994,104]\or[992,121]\or[990,139]\or[987,156]\or[984,173]\or
[981,190]\or[978,207]\or[974,224]\or[970,241]\or[965,258]\or
[961,275]\or[956,292]\or[951,309]\or[945,325]\or[939,342]\or
[933,358]\or[927,374]\or[920,390]\or[913,406]\or[906,422]\or
[898,438]\or[891,453]\or[882,469]\or[874,484]\or[866,499]\or
[857,515]\or[848,529]\or[838,544]\or[829,559]\or[819,573]\or
[809,587]\or[798,601]\or[788,615]\or[777,629]\or[766,642]\or
[754,656]\or[743,669]\or[731,681]\or[719,694]\or[707,707]\or
\else[9999,0]\fi}

\catcode`\[=12

\def\GRAPH(hsize=#1)#2{\hbox to #1\Lengthunit{#2\hss}}

\def\Linewidth#1{\global\linwid*=#1\relax
\global\divide\linwid*10\global\multiply\linwid*\mag
\global\divide\linwid*100\special{em:linewidth \the\linwid*}}

\Linewidth{.4pt}
\def\sm*{\special{em:moveto}}
\def\sl*{\special{em:lineto}}
\let\moveto=\sm*
\let\lineto=\sl*
\newbox\spm*   \newbox\spl*
\setbox\spm*\hbox{\sm*}
\setbox\spl*\hbox{\sl*}

\def\mov#1(#2,#3)#4{\rlap{\L*=#1\Lengthunit
\xL*=#2\L* \yL*=#3\L*
\xL*=\xscale\xL* \yL*=\yscale\yL*
\rx* \the\cos*\xL* \tmp* \the\sin*\yL* \advance\rx*-\tmp*
\ry* \the\cos*\yL* \tmp* \the\sin*\xL* \advance\ry*\tmp*
\kern\rx*\raise\ry*\hbox{#4}}}

\def\rmov*(#1,#2)#3{\rlap{\xL*=#1\yL*=#2\relax
\rx* \the\cos*\xL* \tmp* \the\sin*\yL* \advance\rx*-\tmp*
\ry* \the\cos*\yL* \tmp* \the\sin*\xL* \advance\ry*\tmp*
\kern\rx*\raise\ry*\hbox{#3}}}

\def\lin#1(#2,#3){\rlap{\sm*\mov#1(#2,#3){\sl*}}}

\def\arr*(#1,#2,#3){\rmov*(#1\dd*,#1\dt*){\sm*
\rmov*(#2\dd*,#2\dt*){\rmov*(#3\dt*,-#3\dd*){\sl*}}\sm*
\rmov*(#2\dd*,#2\dt*){\rmov*(-#3\dt*,#3\dd*){\sl*}}}}

\def\arrow#1(#2,#3){\rlap{\lin#1(#2,#3)\mov#1(#2,#3){\relax
\d**=-.012\Lengthunit\dd*=#2\d**\dt*=#3\d**
\arr*(1,10,4)\arr*(3,8,4)\arr*(4.8,4.2,3)}}}

\def\arrlin#1(#2,#3){\rlap{\L*=#1\Lengthunit\L*=.5\L*
\lin#1(#2,#3)\rmov*(#2\L*,#3\L*){\arrow.1(#2,#3)}}}

\def\dasharrow#1(#2,#3){\rlap{{\Lengthunit=0.9\Lengthunit
\dashlin#1(#2,#3)\mov#1(#2,#3){\sm*}}\mov#1(#2,#3){\sl*
\d**=-.012\Lengthunit\dd*=#2\d**\dt*=#3\d**
\arr*(1,10,4)\arr*(3,8,4)\arr*(4.8,4.2,3)}}}

\def\clap#1{\hbox to 0pt{\hss #1\hss}}

\def\ind(#1,#2)#3{\rlap{\L*=.1\Lengthunit
\xL*=#1\L* \yL*=#2\L*
\rx* \the\cos*\xL* \tmp* \the\sin*\yL* \advance\rx*-\tmp*
\ry* \the\cos*\yL* \tmp* \the\sin*\xL* \advance\ry*\tmp*
\kern\rx*\raise\ry*\hbox{\lower2pt\clap{$#3$}}}}

\def\sh*(#1,#2)#3{\rlap{\dm*=\the\n*\d**
\xL*=\xscale\dm* \yL*=\yscale\dm* \xL*=#1\xL* \yL*=#2\yL*
\rx* \the\cos*\xL* \tmp* \the\sin*\yL* \advance\rx*-\tmp*
\ry* \the\cos*\yL* \tmp* \the\sin*\xL* \advance\ry*\tmp*
\kern\rx*\raise\ry*\hbox{#3}}}

\def\calcnum*#1(#2,#3){\a*=1000sp\b*=1000sp\a*=#2\a*\b*=#3\b*
\ifdim\a*<0pt\a*-\a*\fi\ifdim\b*<0pt\b*-\b*\fi
\ifdim\a*>\b*\c*=.96\a*\advance\c*.4\b*
\else\c*=.96\b*\advance\c*.4\a*\fi
\k*\a*\multiply\k*\k*\l*\b*\multiply\l*\l*
\m*\k*\advance\m*\l*\n*\c*\r*\n*\multiply\n*\n*
\dn*\m*\advance\dn*-\n*\divide\dn*2\divide\dn*\r*
\advance\r*\dn*
\c*=\the\Nhalfperiods5sp\c*=#1\c*\ifdim\c*<0pt\c*-\c*\fi
\multiply\c*\r*\N*\c*\divide\N*10000}

\def\dashlin#1(#2,#3){\rlap{\calcnum*#1(#2,#3)\relax
\d**=#1\Lengthunit\ifdim\d**<0pt\d**-\d**\fi
\divide\N*2\multiply\N*2\advance\N*\*one
\divide\d**\N*\sm*\n*\*one\sh*(#2,#3){\sl*}\loop
\advance\n*\*one\sh*(#2,#3){\sm*}\advance\n*\*one
\sh*(#2,#3){\sl*}\ifnum\n*<\N*\repeat}}

\def\dashdotlin#1(#2,#3){\rlap{\calcnum*#1(#2,#3)\relax
\d**=#1\Lengthunit\ifdim\d**<0pt\d**-\d**\fi
\divide\N*2\multiply\N*2\advance\N*1\multiply\N*2\relax
\divide\d**\N*\sm*\n*\*two\sh*(#2,#3){\sl*}\loop
\advance\n*\*one\sh*(#2,#3){\kern-1.48pt\lower.5pt\hbox{\rm.}}\relax
\advance\n*\*one\sh*(#2,#3){\sm*}\advance\n*\*two
\sh*(#2,#3){\sl*}\ifnum\n*<\N*\repeat}}

\def\shl*(#1,#2)#3{\kern#1#3\lower#2#3\hbox{\unhcopy\spl*}}

\def\trianglin#1(#2,#3){\rlap{\toks0={#2}\toks1={#3}\calcnum*#1(#2,#3)\relax
\dd*=.57\Lengthunit\dd*=#1\dd*\divide\dd*\N*
\divide\dd*\*ths \multiply\dd*\magnitude
\d**=#1\Lengthunit\ifdim\d**<0pt\d**-\d**\fi
\multiply\N*2\divide\d**\N*\sm*\n*\*one\loop
\shl**{\dd*}\dd*-\dd*\advance\n*2\relax
\ifnum\n*<\N*\repeat\n*\N*\shl**{0pt}}}

\def\wavelin#1(#2,#3){\rlap{\toks0={#2}\toks1={#3}\calcnum*#1(#2,#3)\relax
\dd*=.23\Lengthunit\dd*=#1\dd*\divide\dd*\N*
\divide\dd*\*ths \multiply\dd*\magnitude
\d**=#1\Lengthunit\ifdim\d**<0pt\d**-\d**\fi
\multiply\N*4\divide\d**\N*\sm*\n*\*one\loop
\shl**{\dd*}\dt*=1.3\dd*\advance\n*\*one
\shl**{\dt*}\advance\n*\*one
\shl**{\dd*}\advance\n*\*two
\dd*-\dd*\ifnum\n*<\N*\repeat\n*\N*\shl**{0pt}}}

\def\w*lin(#1,#2){\rlap{\toks0={#1}\toks1={#2}\d**=\Lengthunit\dd*=-.12\d**
\divide\dd*\*ths \multiply\dd*\magnitude
\N*8\divide\d**\N*\sm*\n*\*one\loop
\shl**{\dd*}\dt*=1.3\dd*\advance\n*\*one
\shl**{\dt*}\advance\n*\*one
\shl**{\dd*}\advance\n*\*one
\shl**{0pt}\dd*-\dd*\advance\n*1\ifnum\n*<\N*\repeat}}

\def\l*arc(#1,#2)[#3][#4]{\rlap{\toks0={#1}\toks1={#2}\d**=\Lengthunit
\dd*=#3.037\d**\dd*=#4\dd*\dt*=#3.049\d**\dt*=#4\dt*\ifdim\d**>10mm\relax
\d**=.25\d**\n*\*one\shl**{-\dd*}\n*\*two\shl**{-\dt*}\n*3\relax
\shl**{-\dd*}\n*4\relax\shl**{0pt}\else
\ifdim\d**>5mm\d**=.5\d**\n*\*one\shl**{-\dt*}\n*\*two
\shl**{0pt}\else\n*\*one\shl**{0pt}\fi\fi}}

\def\d*arc(#1,#2)[#3][#4]{\rlap{\toks0={#1}\toks1={#2}\d**=\Lengthunit
\dd*=#3.037\d**\dd*=#4\dd*\d**=.25\d**\sm*\n*\*one\shl**{-\dd*}\relax
\n*3\relax\sh*(#1,#2){\xL*=\xscale\dd*\yL*=\yscale\dd*
\kern#2\xL*\lower#1\yL*\hbox{\sm*}}\n*4\relax\shl**{0pt}}}

\def\shl**#1{\c*=\the\n*\d**\d*=#1\relax
\a*=\the\toks0\c*\b*=\the\toks1\d*\advance\a*-\b*
\b*=\the\toks1\c*\d*=\the\toks0\d*\advance\b*\d*
\a*=\xscale\a*\b*=\yscale\b*
\rx* \the\cos*\a* \tmp* \the\sin*\b* \advance\rx*-\tmp*
\ry* \the\cos*\b* \tmp* \the\sin*\a* \advance\ry*\tmp*
\raise\ry*\rlap{\kern\rx*\unhcopy\spl*}}

\def\wlin*#1(#2,#3)[#4]{\rlap{\toks0={#2}\toks1={#3}\relax
\c*=#1\l*\c*\c*=.01\Lengthunit\m*\c*\divide\l*\m*
\c*=\the\Nhalfperiods5sp\multiply\c*\l*\N*\c*\divide\N*\*ths
\divide\N*2\multiply\N*2\advance\N*\*one
\dd*=.002\Lengthunit\dd*=#4\dd*\multiply\dd*\l*\divide\dd*\N*
\divide\dd*\*ths \multiply\dd*\magnitude
\d**=#1\multiply\N*4\divide\d**\N*\sm*\n*\*one\loop
\shl**{\dd*}\dt*=1.3\dd*\advance\n*\*one
\shl**{\dt*}\advance\n*\*one
\shl**{\dd*}\advance\n*\*two
\dd*-\dd*\ifnum\n*<\N*\repeat\n*\N*\shl**{0pt}}}

\def\wavebox#1{\setbox0\hbox{#1}\relax
\a*=\wd0\advance\a*14pt\b*=\ht0\advance\b*\dp0\advance\b*14pt\relax
\hbox{\kern9pt\relax
\rmov*(0pt,\ht0){\rmov*(-7pt,7pt){\wlin*\a*(1,0)[+]\wlin*\b*(0,-1)[-]}}\relax
\rmov*(\wd0,-\dp0){\rmov*(7pt,-7pt){\wlin*\a*(-1,0)[+]\wlin*\b*(0,1)[-]}}\relax
\box0\kern9pt}}

\def\rectangle#1(#2,#3){\relax
\lin#1(#2,0)\lin#1(0,#3)\mov#1(0,#3){\lin#1(#2,0)}\mov#1(#2,0){\lin#1(0,#3)}}

\def\dashrectangle#1(#2,#3){\dashlin#1(#2,0)\dashlin#1(0,#3)\relax
\mov#1(0,#3){\dashlin#1(#2,0)}\mov#1(#2,0){\dashlin#1(0,#3)}}

\def\waverectangle#1(#2,#3){\L*=#1\Lengthunit\a*=#2\L*\b*=#3\L*
\ifdim\a*<0pt\a*-\a*\def\x*{-1}\else\def\x*{1}\fi
\ifdim\b*<0pt\b*-\b*\def\y*{-1}\else\def\y*{1}\fi
\wlin*\a*(\x*,0)[-]\wlin*\b*(0,\y*)[+]\relax
\mov#1(0,#3){\wlin*\a*(\x*,0)[+]}\mov#1(#2,0){\wlin*\b*(0,\y*)[-]}}

\def\calcparab*{\ifnum\n*>\m*\k*\N*\advance\k*-\n*\else\k*\n*\fi
\a*=\the\k* sp\a*=10\a*\b*\dm*\advance\b*-\a*\k*\b*
\a*=\the\*ths\b*\divide\a*\l*\multiply\a*\k*
\divide\a*\l*\k*\*ths\r*\a*\advance\k*-\r*\dt*=\the\k*\L*}

\def\arcto#1(#2,#3)[#4]{\rlap{\toks0={#2}\toks1={#3}\calcnum*#1(#2,#3)\relax
\dm*=135sp\dm*=#1\dm*\d**=#1\Lengthunit\ifdim\dm*<0pt\dm*-\dm*\fi
\multiply\dm*\r*\a*=.3\dm*\a*=#4\a*\ifdim\a*<0pt\a*-\a*\fi
\advance\dm*\a*\N*\dm*\divide\N*10000\relax
\divide\N*2\multiply\N*2\advance\N*\*one
\L*=-.25\d**\L*=#4\L*\divide\d**\N*\divide\L*\*ths
\m*\N*\divide\m*2\dm*=\the\m*5sp\l*\dm*\sm*\n*\*one\loop
\calcparab*\shl**{-\dt*}\advance\n*1\ifnum\n*<\N*\repeat}}

\def\arrarcto#1(#2,#3)[#4]{\L*=#1\Lengthunit\L*=.54\L*
\arcto#1(#2,#3)[#4]\rmov*(#2\L*,#3\L*){\d*=.457\L*\d*=#4\d*\d**-\d*
\rmov*(#3\d**,#2\d*){\arrow.02(#2,#3)}}}

\def\dasharcto#1(#2,#3)[#4]{\rlap{\toks0={#2}\toks1={#3}\relax
\calcnum*#1(#2,#3)\dm*=\the\N*5sp\a*=.3\dm*\a*=#4\a*\ifdim\a*<0pt\a*-\a*\fi
\advance\dm*\a*\N*\dm*
\divide\N*20\multiply\N*2\advance\N*1\d**=#1\Lengthunit
\L*=-.25\d**\L*=#4\L*\divide\d**\N*\divide\L*\*ths
\m*\N*\divide\m*2\dm*=\the\m*5sp\l*\dm*
\sm*\n*\*one\loop\calcparab*
\shl**{-\dt*}\advance\n*1\ifnum\n*>\N*\else\calcparab*
\sh*(#2,#3){\xL*=#3\dt* \yL*=#2\dt*
\rx* \the\cos*\xL* \tmp* \the\sin*\yL* \advance\rx*\tmp*
\ry* \the\cos*\yL* \tmp* \the\sin*\xL* \advance\ry*-\tmp*
\kern\rx*\lower\ry*\hbox{\sm*}}\fi
\advance\n*1\ifnum\n*<\N*\repeat}}

\def\*shl*#1{\c*=\the\n*\d**\advance\c*#1\a**\d*\dt*\advance\d*#1\b**
\a*=\the\toks0\c*\b*=\the\toks1\d*\advance\a*-\b*
\b*=\the\toks1\c*\d*=\the\toks0\d*\advance\b*\d*
\rx* \the\cos*\a* \tmp* \the\sin*\b* \advance\rx*-\tmp*
\ry* \the\cos*\b* \tmp* \the\sin*\a* \advance\ry*\tmp*
\raise\ry*\rlap{\kern\rx*\unhcopy\spl*}}

\def\calcnormal*#1{\b**=10000sp\a**\b**\k*\n*\advance\k*-\m*
\multiply\a**\k*\divide\a**\m*\a**=#1\a**\ifdim\a**<0pt\a**-\a**\fi
\ifdim\a**>\b**\d*=.96\a**\advance\d*.4\b**
\else\d*=.96\b**\advance\d*.4\a**\fi
\d*=.01\d*\r*\d*\divide\a**\r*\divide\b**\r*
\ifnum\k*<0\a**-\a**\fi\d*=#1\d*\ifdim\d*<0pt\b**-\b**\fi
\k*\a**\a**=\the\k*\dd*\k*\b**\b**=\the\k*\dd*}

\def\wavearcto#1(#2,#3)[#4]{\rlap{\toks0={#2}\toks1={#3}\relax
\calcnum*#1(#2,#3)\c*=\the\N*5sp\a*=.4\c*\a*=#4\a*\ifdim\a*<0pt\a*-\a*\fi
\advance\c*\a*\N*\c*\divide\N*20\multiply\N*2\advance\N*-1\multiply\N*4\relax
\d**=#1\Lengthunit\dd*=.012\d**
\divide\dd*\*ths \multiply\dd*\magnitude
\ifdim\d**<0pt\d**-\d**\fi\L*=.25\d**
\divide\d**\N*\divide\dd*\N*\L*=#4\L*\divide\L*\*ths
\m*\N*\divide\m*2\dm*=\the\m*0sp\l*\dm*
\sm*\n*\*one\loop\calcnormal*{#4}\calcparab*
\*shl*{1}\advance\n*\*one\calcparab*
\*shl*{1.3}\advance\n*\*one\calcparab*
\*shl*{1}\advance\n*2\dd*-\dd*\ifnum\n*<\N*\repeat\n*\N*\shl**{0pt}}}

\def\triangarcto#1(#2,#3)[#4]{\rlap{\toks0={#2}\toks1={#3}\relax
\calcnum*#1(#2,#3)\c*=\the\N*5sp\a*=.4\c*\a*=#4\a*\ifdim\a*<0pt\a*-\a*\fi
\advance\c*\a*\N*\c*\divide\N*20\multiply\N*2\advance\N*-1\multiply\N*2\relax
\d**=#1\Lengthunit\dd*=.012\d**
\divide\dd*\*ths \multiply\dd*\magnitude
\ifdim\d**<0pt\d**-\d**\fi\L*=.25\d**
\divide\d**\N*\divide\dd*\N*\L*=#4\L*\divide\L*\*ths
\m*\N*\divide\m*2\dm*=\the\m*0sp\l*\dm*
\sm*\n*\*one\loop\calcnormal*{#4}\calcparab*
\*shl*{1}\advance\n*2\dd*-\dd*\ifnum\n*<\N*\repeat\n*\N*\shl**{0pt}}}

\def\hr*#1{\L*=\xscale\Lengthunit\ifnum
\angle**=0\clap{\vrule width#1\L* height.1pt}\else
\L*=#1\L*\L*=.5\L*\rmov*(-\L*,0pt){\sm*}\rmov*(\L*,0pt){\sl*}\fi}

\def\shade#1[#2]{\rlap{\Lengthunit=#1\Lengthunit
\special{em:linewidth .001pt}\relax
\mov(0,#2.05){\hr*{.994}}\mov(0,#2.1){\hr*{.980}}\relax
\mov(0,#2.15){\hr*{.953}}\mov(0,#2.2){\hr*{.916}}\relax
\mov(0,#2.25){\hr*{.867}}\mov(0,#2.3){\hr*{.798}}\relax
\mov(0,#2.35){\hr*{.715}}\mov(0,#2.4){\hr*{.603}}\relax
\mov(0,#2.45){\hr*{.435}}\special{em:linewidth \the\linwid*}}}

\def\dshade#1[#2]{\rlap{\special{em:linewidth .001pt}\relax
\Lengthunit=#1\Lengthunit\if#2-\def\t*{+}\else\def\t*{-}\fi
\mov(0,\t*.025){\relax
\mov(0,#2.05){\hr*{.995}}\mov(0,#2.1){\hr*{.988}}\relax
\mov(0,#2.15){\hr*{.969}}\mov(0,#2.2){\hr*{.937}}\relax
\mov(0,#2.25){\hr*{.893}}\mov(0,#2.3){\hr*{.836}}\relax
\mov(0,#2.35){\hr*{.760}}\mov(0,#2.4){\hr*{.662}}\relax
\mov(0,#2.45){\hr*{.531}}\mov(0,#2.5){\hr*{.320}}\relax
\special{em:linewidth \the\linwid*}}}}

\def\vdot{\rlap{\kern-1.9pt\lower1.8pt\hbox{$\scriptstyle\bullet$}}}
\def\vtimes{\rlap{\kern-3pt\lower1.8pt\hbox{$\scriptstyle\times$}}}
\def\vDot{\rlap{\kern-2.3pt\lower2.7pt\hbox{$\bullet$}}}
\def\vTimes{\rlap{\kern-3.6pt\lower2.4pt\hbox{$\times$}}}

\def\arc(#1)[#2,#3]{{\k*=#2\l*=#3\m*=\l*
\advance\m*-6\ifnum\k*>\l*\relax\else
{\rotate(#2)\mov(#1,0){\sm*}}\loop
\ifnum\k*<\m*\advance\k*5{\rotate(\k*)\mov(#1,0){\sl*}}\repeat
{\rotate(#3)\mov(#1,0){\sl*}}\fi}}

\def\dasharc(#1)[#2,#3]{{\k**=#2\n*=#3\advance\n*-1\advance\n*-\k**
\L*=1000sp\L*#1\L* \multiply\L*\n* \multiply\L*\Nhalfperiods
\divide\L*57\N*\L* \divide\N*2000\ifnum\N*=0\N*1\fi
\r*\n*  \divide\r*\N* \ifnum\r*<2\r*2\fi
\m**\r* \divide\m**2 \l**\r* \advance\l**-\m** \N*\n* \divide\N*\r*
\k**\r* \multiply\k**\N* \dn*\n* \advance\dn*-\k** 
\divide\dn*2\advance\dn*\*one
\r*\l** \divide\r*2\advance\dn*\r* \advance\N*-2\k**#2\relax
\ifnum\l**<6{\rotate(#2)\mov(#1,0){\sm*}}\advance\k**\dn*
{\rotate(\k**)\mov(#1,0){\sl*}}\advance\k**\m**
{\rotate(\k**)\mov(#1,0){\sm*}}\loop
\advance\k**\l**{\rotate(\k**)\mov(#1,0){\sl*}}\advance\k**\m**
{\rotate(\k**)\mov(#1,0){\sm*}}\advance\N*-1\ifnum\N*>0\repeat
{\rotate(#3)\mov(#1,0){\sl*}}\else\advance\k**\dn*
\arc(#1)[#2,\k**]\loop\advance\k**\m** \r*\k**
\advance\k**\l** {\arc(#1)[\r*,\k**]}\relax
\advance\N*-1\ifnum\N*>0\repeat
\advance\k**\m**\arc(#1)[\k**,#3]\fi}}

\def\triangarc#1(#2)[#3,#4]{{\k**=#3\n*=#4\advance\n*-\k**
\L*=1000sp\L*#2\L* \multiply\L*\n* \multiply\L*\Nhalfperiods
\divide\L*57\N*\L* \divide\N*1000\ifnum\N*=0\N*1\fi
\d**=#2\Lengthunit \d*\d** \divide\d*57\multiply\d*\n*
\r*\n*  \divide\r*\N* \ifnum\r*<2\r*2\fi
\m**\r* \divide\m**2 \l**\r* \advance\l**-\m** \N*\n* \divide\N*\r*
\dt*\d* \divide\dt*\N* \dt*.5\dt* \dt*#1\dt*
\divide\dt*1000\multiply\dt*\magnitude
\k**\r* \multiply\k**\N* \dn*\n* \advance\dn*-\k** \divide\dn*2\relax
\r*\l** \divide\r*2\advance\dn*\r* \advance\N*-1\k**#3\relax
{\rotate(#3)\mov(#2,0){\sm*}}\advance\k**\dn*
{\rotate(\k**)\mov(#2,0){\sl*}}\advance\k**-\m**\advance\l**\m**\loop\dt*-\dt*
\d*\d** \advance\d*\dt*
\advance\k**\l**{\rotate(\k**)\rmov*(\d*,0pt){\sl*}}%
\advance\N*-1\ifnum\N*>0\repeat\advance\k**\m**
{\rotate(\k**)\mov(#2,0){\sl*}}{\rotate(#4)\mov(#2,0){\sl*}}}}

\def\wavearc#1(#2)[#3,#4]{{\k**=#3\n*=#4\advance\n*-\k**
\L*=4000sp\L*#2\L* \multiply\L*\n* \multiply\L*\Nhalfperiods
\divide\L*57\N*\L* \divide\N*1000\ifnum\N*=0\N*1\fi
\d**=#2\Lengthunit \d*\d** \divide\d*57\multiply\d*\n*
\r*\n*  \divide\r*\N* \ifnum\r*=0\r*1\fi
\m**\r* \divide\m**2 \l**\r* \advance\l**-\m** \N*\n* \divide\N*\r*
\dt*\d* \divide\dt*\N* \dt*.7\dt* \dt*#1\dt*
\divide\dt*1000\multiply\dt*\magnitude
\k**\r* \multiply\k**\N* \dn*\n* \advance\dn*-\k** \divide\dn*2\relax
\divide\N*4\advance\N*-1\k**#3\relax
{\rotate(#3)\mov(#2,0){\sm*}}\advance\k**\dn*
{\rotate(\k**)\mov(#2,0){\sl*}}\advance\k**-\m**\advance\l**\m**\loop\dt*-\dt*
\d*\d** \advance\d*\dt* \dd*\d** \advance\dd*1.3\dt*
\advance\k**\r*{\rotate(\k**)\rmov*(\d*,0pt){\sl*}}\relax
\advance\k**\r*{\rotate(\k**)\rmov*(\dd*,0pt){\sl*}}\relax
\advance\k**\r*{\rotate(\k**)\rmov*(\d*,0pt){\sl*}}\relax
\advance\k**\r*
\advance\N*-1\ifnum\N*>0\repeat\advance\k**\m**
{\rotate(\k**)\mov(#2,0){\sl*}}{\rotate(#4)\mov(#2,0){\sl*}}}}

\def\gmov*#1(#2,#3)#4{\rlap{\L*=#1\Lengthunit
\xL*=#2\L* \yL*=#3\L*
\rx* \gcos*\xL* \tmp* \gsin*\yL* \advance\rx*-\tmp*
\ry* \gcos*\yL* \tmp* \gsin*\xL* \advance\ry*\tmp*
\rx*=\xscale\rx* \ry*=\yscale\ry*
\xL* \the\cos*\rx* \tmp* \the\sin*\ry* \advance\xL*-\tmp*
\yL* \the\cos*\ry* \tmp* \the\sin*\rx* \advance\yL*\tmp*
\kern\xL*\raise\yL*\hbox{#4}}}

\def\rgmov*(#1,#2)#3{\rlap{\xL*#1\yL*#2\relax
\rx* \gcos*\xL* \tmp* \gsin*\yL* \advance\rx*-\tmp*
\ry* \gcos*\yL* \tmp* \gsin*\xL* \advance\ry*\tmp*
\rx*=\xscale\rx* \ry*=\yscale\ry*
\xL* \the\cos*\rx* \tmp* \the\sin*\ry* \advance\xL*-\tmp*
\yL* \the\cos*\ry* \tmp* \the\sin*\rx* \advance\yL*\tmp*
\kern\xL*\raise\yL*\hbox{#3}}}

\def\Earc(#1)[#2,#3][#4,#5]{{\k*=#2\l*=#3\m*=\l*
\advance\m*-6\ifnum\k*>\l*\relax\else\def\xscale{#4}\def\yscale{#5}\relax
{\angle**0\rotate(#2)}\gmov*(#1,0){\sm*}\loop
\ifnum\k*<\m*\advance\k*5\relax
{\angle**0\rotate(\k*)}\gmov*(#1,0){\sl*}\repeat
{\angle**0\rotate(#3)}\gmov*(#1,0){\sl*}\relax
\def\xscale{1}\def\yscale{1}\fi}}

\def\dashEarc(#1)[#2,#3][#4,#5]{{\k**=#2\n*=#3\advance\n*-1\advance\n*-\k**
\L*=1000sp\L*#1\L* \multiply\L*\n* \multiply\L*\Nhalfperiods
\divide\L*57\N*\L* \divide\N*2000\ifnum\N*=0\N*1\fi
\r*\n*  \divide\r*\N* \ifnum\r*<2\r*2\fi
\m**\r* \divide\m**2 \l**\r* \advance\l**-\m** \N*\n* \divide\N*\r*
\k**\r*\multiply\k**\N* \dn*\n* \advance\dn*-\k** \divide\dn*2\advance\dn*\*one
\r*\l** \divide\r*2\advance\dn*\r* \advance\N*-2\k**#2\relax
\ifnum\l**<6\def\xscale{#4}\def\yscale{#5}\relax
{\angle**0\rotate(#2)}\gmov*(#1,0){\sm*}\advance\k**\dn*
{\angle**0\rotate(\k**)}\gmov*(#1,0){\sl*}\advance\k**\m**
{\angle**0\rotate(\k**)}\gmov*(#1,0){\sm*}\loop
\advance\k**\l**{\angle**0\rotate(\k**)}\gmov*(#1,0){\sl*}\advance\k**\m**
{\angle**0\rotate(\k**)}\gmov*(#1,0){\sm*}\advance\N*-1\ifnum\N*>0\repeat
{\angle**0\rotate(#3)}\gmov*(#1,0){\sl*}\def\xscale{1}\def\yscale{1}\else
\advance\k**\dn* \Earc(#1)[#2,\k**][#4,#5]\loop\advance\k**\m** \r*\k**
\advance\k**\l** {\Earc(#1)[\r*,\k**][#4,#5]}\relax
\advance\N*-1\ifnum\N*>0\repeat
\advance\k**\m**\Earc(#1)[\k**,#3][#4,#5]\fi}}

\def\triangEarc#1(#2)[#3,#4][#5,#6]{{\k**=#3\n*=#4\advance\n*-\k**
\L*=1000sp\L*#2\L* \multiply\L*\n* \multiply\L*\Nhalfperiods
\divide\L*57\N*\L* \divide\N*1000\ifnum\N*=0\N*1\fi
\d**=#2\Lengthunit \d*\d** \divide\d*57\multiply\d*\n*
\r*\n*  \divide\r*\N* \ifnum\r*<2\r*2\fi
\m**\r* \divide\m**2 \l**\r* \advance\l**-\m** \N*\n* \divide\N*\r*
\dt*\d* \divide\dt*\N* \dt*.5\dt* \dt*#1\dt*
\divide\dt*1000\multiply\dt*\magnitude
\k**\r* \multiply\k**\N* \dn*\n* \advance\dn*-\k** \divide\dn*2\relax
\r*\l** \divide\r*2\advance\dn*\r* \advance\N*-1\k**#3\relax
\def\xscale{#5}\def\yscale{#6}\relax
{\angle**0\rotate(#3)}\gmov*(#2,0){\sm*}\advance\k**\dn*
{\angle**0\rotate(\k**)}\gmov*(#2,0){\sl*}\advance\k**-\m**
\advance\l**\m**\loop\dt*-\dt* \d*\d** \advance\d*\dt*
\advance\k**\l**{\angle**0\rotate(\k**)}\rgmov*(\d*,0pt){\sl*}\relax
\advance\N*-1\ifnum\N*>0\repeat\advance\k**\m**
{\angle**0\rotate(\k**)}\gmov*(#2,0){\sl*}\relax
{\angle**0\rotate(#4)}\gmov*(#2,0){\sl*}\def\xscale{1}\def\yscale{1}}}

\def\waveEarc#1(#2)[#3,#4][#5,#6]{{\k**=#3\n*=#4\advance\n*-\k**
\L*=4000sp\L*#2\L* \multiply\L*\n* \multiply\L*\Nhalfperiods
\divide\L*57\N*\L* \divide\N*1000\ifnum\N*=0\N*1\fi
\d**=#2\Lengthunit \d*\d** \divide\d*57\multiply\d*\n*
\r*\n*  \divide\r*\N* \ifnum\r*=0\r*1\fi
\m**\r* \divide\m**2 \l**\r* \advance\l**-\m** \N*\n* \divide\N*\r*
\dt*\d* \divide\dt*\N* \dt*.7\dt* \dt*#1\dt*
\divide\dt*1000\multiply\dt*\magnitude
\k**\r* \multiply\k**\N* \dn*\n* \advance\dn*-\k** \divide\dn*2\relax
\divide\N*4\advance\N*-1\k**#3\def\xscale{#5}\def\yscale{#6}\relax
{\angle**0\rotate(#3)}\gmov*(#2,0){\sm*}\advance\k**\dn*
{\angle**0\rotate(\k**)}\gmov*(#2,0){\sl*}\advance\k**-\m**
\advance\l**\m**\loop\dt*-\dt*
\d*\d** \advance\d*\dt* \dd*\d** \advance\dd*1.3\dt*
\advance\k**\r*{\angle**0\rotate(\k**)}\rgmov*(\d*,0pt){\sl*}\relax
\advance\k**\r*{\angle**0\rotate(\k**)}\rgmov*(\dd*,0pt){\sl*}\relax
\advance\k**\r*{\angle**0\rotate(\k**)}\rgmov*(\d*,0pt){\sl*}\relax
\advance\k**\r*
\advance\N*-1\ifnum\N*>0\repeat\advance\k**\m**
{\angle**0\rotate(\k**)}\gmov*(#2,0){\sl*}\relax
{\angle**0\rotate(#4)}\gmov*(#2,0){\sl*}\def\xscale{1}\def\yscale{1}}}

\newcount\CatcodeOfAtSign
\CatcodeOfAtSign=\the\catcode`\@
\catcode`\@=11
\def\@arc#1[#2][#3]{\rlap{\Lengthunit=#1\Lengthunit
\sm*\l*arc(#2.1914,#3.0381)[#2][#3]\relax
\mov(#2.1914,#3.0381){\l*arc(#2.1622,#3.1084)[#2][#3]}\relax
\mov(#2.3536,#3.1465){\l*arc(#2.1084,#3.1622)[#2][#3]}\relax
\mov(#2.4619,#3.3086){\l*arc(#2.0381,#3.1914)[#2][#3]}}}

\def\dash@arc#1[#2][#3]{\rlap{\Lengthunit=#1\Lengthunit
\d*arc(#2.1914,#3.0381)[#2][#3]\relax
\mov(#2.1914,#3.0381){\d*arc(#2.1622,#3.1084)[#2][#3]}\relax
\mov(#2.3536,#3.1465){\d*arc(#2.1084,#3.1622)[#2][#3]}\relax
\mov(#2.4619,#3.3086){\d*arc(#2.0381,#3.1914)[#2][#3]}}}

\def\wave@arc#1[#2][#3]{\rlap{\Lengthunit=#1\Lengthunit
\w*lin(#2.1914,#3.0381)\relax
\mov(#2.1914,#3.0381){\w*lin(#2.1622,#3.1084)}\relax
\mov(#2.3536,#3.1465){\w*lin(#2.1084,#3.1622)}\relax
\mov(#2.4619,#3.3086){\w*lin(#2.0381,#3.1914)}}}

\def\bezier#1(#2,#3)(#4,#5)(#6,#7){\N*#1\l*\N* \advance\l*\*one
\d* #4\Lengthunit \advance\d* -#2\Lengthunit \multiply\d* \*two
\b* #6\Lengthunit \advance\b* -#2\Lengthunit
\advance\b*-\d* \divide\b*\N*
\d** #5\Lengthunit \advance\d** -#3\Lengthunit \multiply\d** \*two
\b** #7\Lengthunit \advance\b** -#3\Lengthunit
\advance\b** -\d** \divide\b**\N*
\mov(#2,#3){\sm*{\loop\ifnum\m*<\l*
\a*\m*\b* \advance\a*\d* \divide\a*\N* \multiply\a*\m*
\a**\m*\b** \advance\a**\d** \divide\a**\N* \multiply\a**\m*
\rmov*(\a*,\a**){\unhcopy\spl*}\advance\m*\*one\repeat}}}

\catcode`\*=12

\newcount\n@ast
\def\n@ast@#1{\n@ast0\relax\get@ast@#1\end}
\def\get@ast@#1{\ifx#1\end\let\next\relax\else
\ifx#1*\advance\n@ast1\fi\let\next\get@ast@\fi\next}

\newif\if@up \newif\if@dwn
\def\up@down@#1{\@upfalse\@dwnfalse
\if#1u\@uptrue\fi\if#1U\@uptrue\fi\if#1+\@uptrue\fi
\if#1d\@dwntrue\fi\if#1D\@dwntrue\fi\if#1-\@dwntrue\fi}

\def\halfcirc#1(#2)[#3]{{\Lengthunit=#2\Lengthunit\up@down@{#3}\relax
\if@up\mov(0,.5){\@arc[-][-]\@arc[+][-]}\fi
\if@dwn\mov(0,-.5){\@arc[-][+]\@arc[+][+]}\fi
\def\lft{\mov(0,.5){\@arc[-][-]}\mov(0,-.5){\@arc[-][+]}}\relax
\def\rght{\mov(0,.5){\@arc[+][-]}\mov(0,-.5){\@arc[+][+]}}\relax
\if#3l\lft\fi\if#3L\lft\fi\if#3r\rght\fi\if#3R\rght\fi
\n@ast@{#1}\relax
\ifnum\n@ast>0\if@up\shade[+]\fi\if@dwn\shade[-]\fi\fi
\ifnum\n@ast>1\if@up\dshade[+]\fi\if@dwn\dshade[-]\fi\fi}}

\def\halfdashcirc(#1)[#2]{{\Lengthunit=#1\Lengthunit\up@down@{#2}\relax
\if@up\mov(0,.5){\dash@arc[-][-]\dash@arc[+][-]}\fi
\if@dwn\mov(0,-.5){\dash@arc[-][+]\dash@arc[+][+]}\fi
\def\lft{\mov(0,.5){\dash@arc[-][-]}\mov(0,-.5){\dash@arc[-][+]}}\relax
\def\rght{\mov(0,.5){\dash@arc[+][-]}\mov(0,-.5){\dash@arc[+][+]}}\relax
\if#2l\lft\fi\if#2L\lft\fi\if#2r\rght\fi\if#2R\rght\fi}}

\def\halfwavecirc(#1)[#2]{{\Lengthunit=#1\Lengthunit\up@down@{#2}\relax
\if@up\mov(0,.5){\wave@arc[-][-]\wave@arc[+][-]}\fi
\if@dwn\mov(0,-.5){\wave@arc[-][+]\wave@arc[+][+]}\fi
\def\lft{\mov(0,.5){\wave@arc[-][-]}\mov(0,-.5){\wave@arc[-][+]}}\relax
\def\rght{\mov(0,.5){\wave@arc[+][-]}\mov(0,-.5){\wave@arc[+][+]}}\relax
\if#2l\lft\fi\if#2L\lft\fi\if#2r\rght\fi\if#2R\rght\fi}}

\catcode`\*=11

\def\Circle#1(#2){\halfcirc#1(#2)[u]\halfcirc#1(#2)[d]\n@ast@{#1}\relax
\ifnum\n@ast>0\L*=\xscale\Lengthunit
\ifnum\angle**=0\clap{\vrule width#2\L* height.1pt}\else
\L*=#2\L*\L*=.5\L*\special{em:linewidth .001pt}\relax
\rmov*(-\L*,0pt){\sm*}\rmov*(\L*,0pt){\sl*}\relax
\special{em:linewidth \the\linwid*}\fi\fi}

\catcode`\*=12

\def\wavecirc(#1){\halfwavecirc(#1)[u]\halfwavecirc(#1)[d]}

\def\dashcirc(#1){\halfdashcirc(#1)[u]\halfdashcirc(#1)[d]}

\def\xscale{1}
\def\yscale{1}

\def\Ellipse#1(#2)[#3,#4]{\def\xscale{#3}\def\yscale{#4}\relax
\Circle#1(#2)\def\xscale{1}\def\yscale{1}}

\def\dashEllipse(#1)[#2,#3]{\def\xscale{#2}\def\yscale{#3}\relax
\dashcirc(#1)\def\xscale{1}\def\yscale{1}}

\def\waveEllipse(#1)[#2,#3]{\def\xscale{#2}\def\yscale{#3}\relax
\wavecirc(#1)\def\xscale{1}\def\yscale{1}}

\def\halfEllipse#1(#2)[#3][#4,#5]{\def\xscale{#4}\def\yscale{#5}\relax
\halfcirc#1(#2)[#3]\def\xscale{1}\def\yscale{1}}

\def\halfdashEllipse(#1)[#2][#3,#4]{\def\xscale{#3}\def\yscale{#4}\relax
\halfdashcirc(#1)[#2]\def\xscale{1}\def\yscale{1}}

\def\halfwaveEllipse(#1)[#2][#3,#4]{\def\xscale{#3}\def\yscale{#4}\relax
\halfwavecirc(#1)[#2]\def\xscale{1}\def\yscale{1}}

\catcode`\@=\the\CatcodeOfAtSign

\section{Introduction}

The decay and production processes of the bound states with heavy
quarks are investigated with greater intensity in the last years.
The research aims of many experiments (ALEPH, DELPHI, SLD, CLEO, Belle, SELEX, 
LHC-b)
are directed on the growth of experimental accuracy in the derivation of the
static characteristics of heavy hadrons, their production
and decay rates in different reactions \cite{BFY,RunII,Z00,QWG}. The
production of heavy mesons and baryons via heavy quark
fragmentation in the $e^+e^-$ - annihilation represents one of the
possible mechanisms for the formation of heavy hadrons with two
heavy quarks. The fragmentation cross sections for the production
of heavy hadrons can be calculated in an analytical form using the
factorization hypothesis. The heavy quark production amplitude can be
calculated on the basis of perturbative QCD. The characteristic
quark virtualities of heavy quarks in the hard production are
of the order of their masses while the quark virtualities in the bound
state are much less than their masses due to the nonrelativistic
motion. So, the total amplitude can be represented as a convolution
of the hard transition amplitude with a nonperturbative factor (the
wave function) determining the transition of free heavy quarks
into a bound state \cite{LSG}. The fragmentation mechanism was used for the
study of the production processes of heavy mesons and baryons in
$e^+e^-$ annihilation in Refs.\cite{B1,CC,B2,B3,UFN1,UFN2,MS1,MS2}
(a more complete list of references can be found in 
Refs.\cite{RunII,QWG,UFN2}).
The growth of theoretical accuracy for the calculation of
corresponding production cross sections can be reached in two
ways. Firstly, it is necessary to take into account radiative
corrections to the perturbative amplitude describing the production
of free heavy quarks via heavy quark fragmentation. Secondly, we
must consistently consider the relativistic corrections in the
fragmentation amplitude connected with the relative motion of
heavy quarks forming heavy hadron. From the point of view of NRQCD
both effects are caused by the matrix elements as a function of the
typical heavy quark velocity in the bound state rest frame of orders
$O(v_Q)$ and $O(v_Q^2)$, respectively \cite{B3}.
The experimental data indicate that
the calculations of different production probabilities for heavy
quarkonium and double heavy baryons should be improved by
a systematic account of relativistic corrections. Such effects
as the relative motion of heavy quarks forming heavy quarkonia
and diquarks, the diquark structure effects in the calculation of
the fragmentation functions of heavy diquarks should be
considered. The role of relativistic effects was studied already
in the processes of c-quark fragmentation into $J/\Psi$, $\eta_c$
in Ref.\cite{Bashir} on the basis of the Bethe-Salpeter approach, the
gluon fragmentation into S-wave quarkonium in Ref.\cite{Bodwin}
and in the inclusive production of polarized $J/\Psi$ from
$b$-quark decay in Ref.\cite{Ma}. The consideration of the intrinsic motion of 
quarks forming the heavy measons can explain the discreapancy between 
theoretical predictions and experimental data for the cross sections of the 
process $e^+e^-\to\Psi\eta_c$ \cite{BC,BLL}.
The aim of the present work is to get a 
systematically improved description of the relativistic effects in the 
processes of the heavy quark fragmentation in the quasipotential approach
\cite{savrin}.
Our goal also consists
in the calculation of the relativistic corrections in
heavy quark $b$ and $c$ fragmentation functions into pseudoscalar
and vector heavy mesons $(Q_1\bar Q_2)$ on the basis of the
relativistic quark model used earlier in the calculation of mass
spectra of heavy mesons and baryons and their decay rates in
different reactions \cite{rqm1,rqm2,rqm3}. In particular, we investigate
double distribution probabilities for the heavy quark
fragmentation over longitudinal meson momentum $z$ and transverse
meson momentum $p_T$. Analytical expressions for the
fragmentation probabilities as functions of transverse
momentum of heavy mesons $(Q_1\bar Q_2)$ are obtained.

\begin{figure}[htbp]
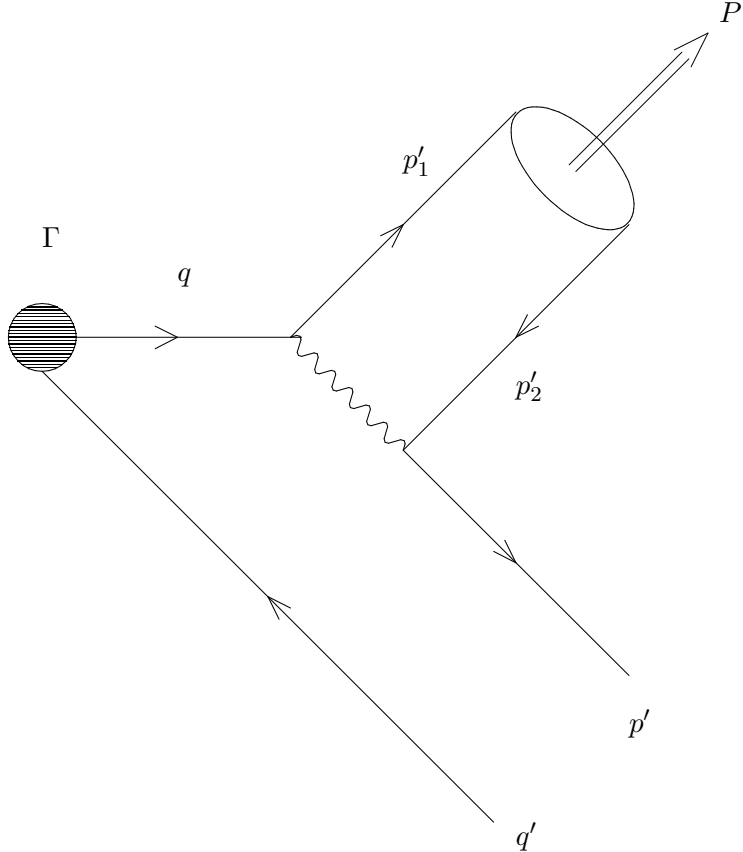

\magnitude=2000
\GRAPH(hsize=15){
\mov(7.03,1.47){\lin(1,1)}%
\mov(6.97,1.53){\lin(1,1)}%
\mov(8.3,2.8){$P$}%
\mov(8.2,2.7){\lin(-0.3,-0.15)}%
\mov(8.2,2.7){\lin(-0.15,-0.3)}%
\mov(2.6,0){\lin(2,0)}%
\mov(2.3,0){\Circle*(0.6)}%
\mov(2.3,0.8){$\Gamma$}%
\mov(4.5,0){\lin(2,2)}%
\mov(4.5,0){\wavelin(1,-1)}%
\mov(5.5,-1){\lin(2,2)}%
\mov(5.5,-1){\lin(2,-2)}%
\mov(2.3,-0.3){\lin(4,-4)}%
\mov(4.3,-2.3){\lin(0.2,-0.1)}%
\mov(4.3,-2.3){\lin(0.1,-0.2)}%
\mov(6.5,-4.5){$q'$}%
\mov(3.5,0){\lin(-0.2,-0.1)}%
\mov(3.5,0){\lin(-0.2,0.1)}%
\mov(7.,1.5){\rotate(45)\Ellipse(1.38)[0.5,1]}%
\mov(5.5,1){\lin(-0.2,-0.1)}%
\mov(5.5,1){\lin(-0.1,-0.2)}%
\mov(3.5,0.5){$q$}%
\mov(6.5,0){\lin(0.2,0.1)}%
\mov(6.5,0){\lin(0.1,0.2)}%
\mov(5.5,1.5){$p'_1$}%
\mov(6.5,-0.5){$p'_2$}%
\mov(7.5,-3.5){$p'$}%
\mov(6.5,-2){\lin(-0.1,0.2)}%
\mov(6.5,-2){\lin(-0.2,0.1)}%
}
\vspace{3mm}
\caption{The Feynman diagram for the fragmentation of a heavy quark $Q_1$ with 
a four-momentum $q$ to a heavy meson $(Q_1\bar Q_2)$ with a four-momentum $P$. 
$\Gamma$ is the vertex function
determining the production of the quark-antiquark pair in the $Z^0$ decay.}
\end{figure}

\section{General formalism}

The heavy meson production through the heavy quark fragmentation is shown in 
Fig.1. On the first stage $Z^0$ boson decays into a quark-antiquark pair
with four-momenta $q$ and $q'$ respectively. After that one heavy quark with 
the four-momentum $q$ fragments to the heavy quarkonium.
In the quasipotential approach the
invariant transition amplitude of a heavy quark $b$ or $c$ 
into a heavy meson can be expressed as a simple convolution of a
perturbative production amplitude $T(p'_1,p'_2,p',q')$
of free quarks and
the quasipotential wave function of the bound state $(Q_1\bar
Q_2)$ $\Psi_P({\bf p})$ \cite{savrin,faustov1973}:
\be
M(q,P,p',q')=\int\frac{d{\bf p}}{(2\pi)^3}\bar\Psi_P({\bf 
p})T(p'_1,p'_2,p',q'),
\ee
where four-momenta of fragmenting quarks $(b,c)$ and spectator antiquarks
$(\bar b, \bar c)$ forming the heavy meson are defined as follows:
\begin{equation}
p'_1=\eta_1P+p,~~~p'_2=\eta_2P-p,
\end{equation}
$p'$ is the four-momentum of a free spectator quark $b$ or $c$ and $P$
is the four-momentum of the heavy meson. The coefficients $\eta_{1,2}$ in
the definition (2) are taken in such a way that the following orthogonality
condition is fulfilled:
\begin{equation}
(p\cdot P)=0,~~~\eta_{1,2}=\frac{M^2-m_{2,1}^2+m_{1,2}^2}{2M^2},
\end{equation}
$M=(m_1+m_2+W)$ is the bound state mass.

The transition of the pair of a heavy quark and antiquark into color-singlet 
mesons
can be envisioned as a complicated process in which the colors
and spins of the heavy quark and antiquark play an important role. Different
color-spin nonperturbative factors entering the amplitude $T(p_1,p_2,p',q')$
control the production of the heavy quark bound states. In this process the 
gluon virtuality $k^2\gg \Lambda^2_{QCD}$ and the strong coupling constant 
$\alpha_s(k^2)\ll 1$. Then the hard part of the
fragmentation amplitude (1) in the leading order over $\alpha_s$ takes the 
form:
\begin{equation}
T(p'_1,p'_2,p',q')=\frac{4\alpha_s}{3\sqrt{3}}\frac{D_{\lambda\sigma}(k)}{(s-m_1^2)}
\bar u_1(p'_1)\gamma_\lambda(\hat q+m_1)\Gamma v_1(q')\bar u_2(p')
\gamma_\sigma v_2(p'_2),
\end{equation}
where $\Gamma$ is the vertex function for the transition of the $Z^0$ boson 
into the quark-antiquark pair;
the gluon propagator is taken in the axial gauge with four-vector 
$n=(1,0,0,-1)$:
\begin{equation}
D_{\lambda\sigma}(k)=\frac{1}{k^2+i\epsilon}\left[-g_{\lambda\sigma}+
\frac{k_\sigma n_\lambda+k_\lambda n_\sigma}{k\cdot n}\right],
\end{equation}
$s=q^2$, $k=(q-\eta_1P-p)$=$(\eta_2P-p+p')$ is the gluon four
momentum. The color factor
$(T^a)_{il}(T^a)_{mj}\delta_{ij}/\sqrt{3}$ =
$4\delta_{ml}/3\sqrt{3}$ was already extracted in the amplitude
(4). The transformation law of the bound state wave functions
from the rest frame to the moving one with four-momenta $P$ is
given by  \cite{rqm2,rqm3,faustov1973}
\begin{equation}
\Psi_{P}^{\rho\omega}({\bf p})=D_1^{1/2,~\rho\alpha}(R^W_{L_{P}})
D_2^{1/2,~\omega\beta}(R^W_{L_{P}})\Psi_{0}^{\alpha\beta}({\bf
p}),
\end{equation}
\begin{displaymath}
\bar\Psi_{P}^{\lambda\sigma}({\bf p})
=\bar\Psi^{\varepsilon\tau}_{0}({\bf p})D_1^{+~1/2,~\varepsilon
\lambda}(R^W_{L_{P}})D_2^{+~1/2,~\tau\sigma}(R^W_{L_{P}}),
\end{displaymath}
where $R^W$ is the Wigner rotation, $L_{P}$ is the Lorentz boost
from the meson rest frame to a moving one, and
the rotation matrix $D^{1/2}(R)$ is defined by
\begin{equation}
{1 \ \ \,0\choose 0 \ \ \,1}D^{1/2}_{1,2}(R^W_{L_{P}})=
S^{-1}({\bf p}_{1,2})S({\bf P})S({\bf p}),
\end{equation}
where
$$ S({\bf p})=\sqrt{\frac{\epsilon(p)+m}{2m}}\left(1+\frac{(\bm{\alpha}
{\bf p})} {\epsilon(p)+m}\right) $$ is the usual Lorentz transformation
matrix of the four-spinor. For further transformations of the amplitude
(4) the following relations are useful:
\begin{equation}
S_{\alpha\beta}(\Lambda)u^\lambda_\beta(p)=\sum_{\sigma=\pm 1/2}
u^{\sigma}_\alpha(\Lambda p)D^{1/2}_{\sigma\lambda}(R^W_{\Lambda p}),
\end{equation}
\begin{displaymath}
\bar u^\lambda_\beta(p)S^{-1}_{\beta\alpha}(\Lambda)=\sum_{\sigma=\pm 1/2}
D^{+~1/2}_{\lambda\sigma}(R^W_{\Lambda p})\bar u^\sigma_\alpha(\Lambda p).
\end{displaymath}
Using also the transformation law of the Dirac bispinors to the rest frame
\begin{eqnarray}
\bar u_1({\bf p})=\bar u_1(0)\frac{(\hat
p_1+m_1)}{\sqrt{2\epsilon_1({\bf p}) (\epsilon_1({\bf
p})+m_1)}},~~p_1=(\epsilon_1,{\bf p}),\cr\cr v_2(-{\bf
p})=\frac{(\hat p_2-m_2)}{\sqrt{2\epsilon_2({\bf
p})(\epsilon_2({\bf p})+ m_2)}}v_2(0),~~p_2=(\epsilon_2,-{\bf p}),
\end{eqnarray}
we can introduce the projection operators $\hat\Pi^{P,V}$ onto the states 
$(Q_1\bar Q_2)$
in the meson with total spin 0 and 1 as follows:
\begin{equation}
\hat\Pi^{P,V}=[v_2(0)\bar 
u_1(0)]_{S=0,1}=\gamma_5(\hat\epsilon^\ast)\frac{1+\gamma^0}
{2\sqrt{2}}.
\end{equation}
As a result the heavy quark $b(c)$ fragmentation amplitude into the mesons
$(b\bar c)$, $(b\bar b)$ or $(c\bar c)$ takes the form:
\begin{equation}
\label{basic}
M(q,P,p',q')=\frac{2\alpha_s\sqrt{2M}}{3\sqrt{6}}\int\frac{D_{\lambda\sigma}(k)}{(s-m_1^2)}
\frac{d{\bf p}}{(2\pi)^3}\bar\Psi_0({\bf p})\bar
u_2(p')\gamma_\sigma \frac{(\hat{\tilde
p}_2-m_2)}{\sqrt{2\epsilon_2({\bf p})(\epsilon({\bf p})+
m_2)}}\times
\end{equation}
\begin{displaymath}
\times\hat{\tilde\epsilon}^\ast(v)(\hat v+1)\frac{(\hat{\tilde p}_1+m_1)}
{\sqrt{2\epsilon_1({\bf p})(\epsilon_1({\bf
p})+m_1)}}\gamma_\lambda(\hat q+ m_1)\Gamma_\alpha v_1(q'),
\end{displaymath}
where the four-vectors $\tilde\epsilon$, $\tilde p_{1,2}$ are given by:
\begin{equation}
\tilde\epsilon=L_P(0,{\mathstrut\bm\epsilon})=\left({\mathstrut\bm\epsilon}
{\bf v},{\mathstrut\bm\epsilon}+\frac{({\mathstrut\bm\epsilon}{\bf
v}){\bf v}} {1+v^0}\right),
\end{equation}
\begin{displaymath}
\hat{\tilde p}_{1,2}=S(L_P)\hat
p_{1,2}S^{-1}(L_P),~~S(L_P)(1\pm\gamma^0)S^{-1}(L_P)= \pm(\hat
v\pm 1), ~~\hat v=\frac{\hat P}{M}.
\end{displaymath}

Transforming the bispinor contractions in the numerator of the expression (11)
we can find the following expression for the heavy quark fragmentation
amplitude including the effects of relative motion of the heavy quarks:
\begin{equation}
M(q,P,p',q')=\frac{2\alpha_s\sqrt{2M}}{3\sqrt{6}}\int \frac{\bar\Psi_0({\bf
p})}{\sqrt{\frac{\epsilon_1({\bf p})}{m_1} \frac{(\epsilon_1({\bf
p})+m_1)}{2m_1}}\sqrt{\frac{\epsilon_2 ({\bf
p})}{m_2}\frac{(\epsilon_2({\bf p})+m_2)}{2m_2}}}
\frac{D_{\lambda\sigma}(k)}{(s-m_1^2)}\frac{d{\bf p}}
{(2\pi)^3}\times
\end{equation}
\begin{displaymath}
\times\bar u_2(p')\gamma_\sigma\left[\frac{\hat v-1}{2}+\hat v\frac{{\bf 
p}^2}{2m_2(\epsilon_2+m_2)}
-\frac{\hat{\tilde
p}}{2m_2}\right]\hat{\tilde\epsilon}^\ast(v)(\hat v+1)\times
\end{displaymath}
\begin{displaymath}
\times\left[\frac{\hat
v+1}{2}-\hat v\frac{{\bf p}^2}{2m_1(\epsilon_1+m_1)}+ \frac{\hat{\tilde
p}}{2m_1}\right]\gamma_\lambda(\hat q+m_1)\Gamma_\alpha
v_1(q').
\end{displaymath}

The fragmentation amplitude (13) keeps at least two sources of relativistic 
corrections. The corrections of the first group appear from the 
quark-antiquark interaction operator. They can be taken into account by means 
of the numerical solution of the Schroedinger equation with the relevant 
potential. The second part of these corrections is determined by several 
functions 
depending on the momenta of the relative motion of quarks ${\bf p}$. 
In the limit of zero relative momentum ${\bf p}$ the amplitude $M(q,P,p',q')$
was studied in Ref.\cite{B1,CC,B2}. At last, there exist the one-loop 
corrections to the fragmentation amplitude which we have not considered here.

Heavy quarkonium can be characterized by the hard momentum scale $m$ (the mass 
m of the heavy quarks), the soft momentum scale $mv_Q$ and the ultrasoft 
momentum 
scale $mv_Q^2$.
We assume that the heavy quarkonium is a nonrelativistic system. This implies 
that the ratio ${\bf p}^2/m^2\sim v_Q^2\ll 1$. So,
we introduce the expansion of all factors in Eq.(13) over relative momentum 
${\bf p}$ up to terms of the second order:
\begin{equation}
\frac{1}{k^2}=\frac{1}{k_0^2}+\frac{1}{k_0^4}[2qp-p^2]+\frac{4}{k_0^6}(qp)^2,
~~k_0^2=\eta_2 s+\eta_1m_2^2-\eta_1\eta_2M^2,
\end{equation}
\begin{displaymath}
\frac{1}{kn}=\frac{1}{(qn-\eta_1Pn)}+\frac{pn}{(qn-\eta_1Pn)^2}+
\frac{(pn)^2}{(qn-\eta_1Pn)^3},
\end{displaymath}
\begin{displaymath}
(\hat{\tilde p}_1+m_1)=m_1(\hat v+1)+\hat v\frac{{\bf p}^2}{2m_1}+\hat{\tilde 
p},
~~(\hat{\tilde p}_2-m_2)=m_2(\hat v-1)+\hat v\frac{{\bf p}^2}{2m_2}-
\hat{\tilde p},
\end{displaymath}
\begin{displaymath}
\tilde p=L_P(0,{\bf p})=\left({\bf p v},{\bf p}+\frac{{\bf
v}({\bf p}{\bf v})}{v^0+1}\right).
\end{displaymath}
Substituting the expansions (14) into Eq.(13) we obtain:
\begin{equation}
M(q,P,p',q')=\frac{2\alpha_s\sqrt{2M}}{3\sqrt{6}}\int \bar\Psi_0({\bf p})
\left[1-\frac{3}{8}\frac{{\bf p}^2}{8m_1^2}-\frac{3}{8}\frac{{\bf 
p}^2}{8m_2^2}\right]
\frac{1}{(s-m_1^2)}\frac{d{\bf p}}
{(2\pi)^3}\times
\end{equation}
\begin{displaymath}
\times\bar u_2(p')\gamma_\sigma\left[\frac{\hat v-1}{2}+\hat
v\frac{{\bf p}^2}{4m_2^2} -\frac{\hat{\tilde
p}}{2m_2}\right]\hat{\tilde\epsilon}^\ast(v)(\hat v+1)\times
\end{displaymath}
\begin{displaymath}
\times\left[\frac{\hat
v+1}{2}+\hat v\frac{{\bf p}^2}{4m_1^2}+ \frac{\hat{\tilde
p}}{2m_1}\right]\gamma_\lambda(\hat q+m_1)\Gamma_\alpha
v_1(q')\times
\end{displaymath}
\begin{displaymath}
\times\left[\frac{1}{k_0^2}+\frac{1}{k_0^4}[2qp-p^2]+\frac{4}{k_0^6}(qp)^2\right]
\Biggl\{-g_{\lambda\sigma}+(k_\sigma n_\lambda+k_\lambda n_\sigma)\times
\end{displaymath}
\begin{displaymath}
\times\left[\frac{1}{(qn-\eta_1Pn)}+
\frac{pn}{(qn-\eta_1Pn)^2}+\frac{(pn)^2}{(qn-\eta_1Pn)^3}\right]\Biggr\}.
\end{displaymath}

Let us emphasize that for the system of two heavy quarks relative motion
corrections
entering in the gluon propagator or heavy quark propagators have the same
order $O(v_Q^2)$ contrary to the system including heavy and light quarks.
In the last case leading order relativistic corrections are determined
by the relativistic factors belonging to the light quark but other terms 
contain an
additional small ratio $m_q/m_Q$. The obtained relation (15) which has the 
form of a three dimensional integral in the 
momentum space is valid when the integration is restricted to the soft 
momentum 
region, where the wave function has a significant support. 
Otherwise it would diverge at high momenta. Moreover, our aim consists in
preserving here only the terms of the second order over $|{\bf p}|/m$ omitting 
corrections of higher order.

\section{Heavy quark fragmentation functions into P- and V-mesons}

We use the NRQCD factorization approach to the calculation of the fragmentation
reactions which was developed in Refs.\cite{B1,B2}.
The fragmentation function of the heavy quark $Q_1$ to produce $^1S_0$ or
$^3S_1$ $(Q_1\bar Q_2)$ meson states is determined by the following expression:
\begin{equation}
D_{Q_1\to V(Q_1\bar Q_2)}(z)=\frac{1}{16\pi^2}\int 
ds\cdot\theta\left(s-\frac{M^2}
{z}-\frac{m_2^2}{1-z}\right)\lim_{q_0\to\infty}\frac{|M|^2}{|M_0|^2},
\end{equation}
where $q_0$ is the energy of the fragmentating quark: $q_\mu=(q_0,0,0,
\sqrt{q_0^2-s})$; $M_0=\bar u_1(q)\Gamma v_1(q')$ is the amplitude of
free quark $Q_1$ production on the mass shell, $z$ is the meson longitudinal
momentum fraction relative to the fragmenting heavy quark. Let us consider the
fragmentation production of the vector meson. Omitting the momentum of the 
relative motion
of heavy quarks ${\bf p}$ in Eq.(14) we obtain the fragmentation
amplitude which contains the leading order contribution and the correction
due to the quark bound state energy $W$ ($M=m_1+m_2+W$):
\begin{equation}
M_1=\frac{2\sqrt{2M}\alpha_s\bar\Psi(0)}{3\sqrt{6}}\frac{1}{(s-m_1^2)(\eta_2s+
\eta_1m_2^2-\eta_1\eta_2M^2)}\times
\end{equation}
\begin{displaymath}
\times\Bigl[\bar u_2(p')2\hat{\tilde\epsilon}^\ast(v)(\hat
q+m_1)\Gamma_\alpha v_1(q')+
\frac{(s+\eta_2m_1M-\eta_1M^2)}{(n q-\eta_1nP)}\bar u_2(p')\hat n
\hat{\tilde\epsilon}^\ast(v) (\hat v+1)\Gamma_\alpha v_1(q')+
\end{displaymath}
\begin{displaymath}
+\frac{(m_1-\eta_1M)}{(n q-\eta_1nP)}\bar u_2(p')\hat
n\hat{\tilde\epsilon}^\ast(v) (\hat v+1)\hat p'\Gamma_\alpha v_1(q')+
\frac{(m_2-\eta_2M)}{(n q-\eta_1nP)}\bar u_2(p')\hat
n\hat{\tilde\epsilon}^\ast(v) (\hat v+1)(\hat q+m_1)\Gamma_\alpha v_1
(q')\Bigr].
\end{displaymath}
Taking into account linear terms in the binding energy $W$ in the expansion
of Eq.(16) we next perform an averaging and summation over the meson spins in 
initial and
final states in the square modulus $|M|^2$
\begin{equation}
\frac{1}{3}\sum_{spin}\tilde\epsilon^\ast_\alpha(v)\tilde\epsilon_\beta(v)=
\frac{1}{3}\left(-g_{\alpha\beta}+v_\alpha v_\beta\right),
\end{equation}
and then consider the limit $q_0\to\infty$ in the obtained expression. In this
limit $\hat P$ and $\hat q$ have the order of the $M_Z$ mass and the 
coefficients
in corresponding expressions are determined by the heavy quark masses
of the fragmenting quark $m_1$ and the spectator quark $m_2$.
In the leading order we can substitute $P=zq$, and the trace in $|M|^2$ over
the Dirac indices is proportional to $Tr(\Gamma_\alpha\hat{q'}\Gamma_\beta
\hat{q'})$. It will disappear in the ratio $|M|^2/|M_0|^2$. Then the
fragmentation probability (16) can be written as a sum of two terms:
\begin{equation}
D_{Q_1\to V(Q_1\bar 
Q_2)}(z)=\frac{8\alpha_s^2|\Psi(0)|^2}{27m_2^3}\frac{rz(1-z)^2}
{[1-(1-r)z]^6}(v_0+v_1),
\end{equation}
\begin{equation}
v_0=2-2(3-2r)z+3(3-2r+4r^2)z^2-2(1-r)(4-r+2r^2)z^3+(1-r)^2(3-2r+2r^2)z^4,
\end{equation}
\begin{equation}
v_1=\frac{W}{3m_2[1-(1-r)z]^2}\Bigl[-12+6r+(60-84r+48r^2)z+(-138+259r-
296r^2+102r^3)z^2+
\end{equation}
\begin{displaymath}
+(192-420r+552r^2-536r^3+330r^4)z^3+(-168+446r-614r^2+746r^3-612r^4+232r^5)
z^4+
\end{displaymath}
\begin{displaymath}
+(84-274r+476r^2-650r^3+574r^4-272r^5+62r^6)z^5+(-18+57r-126r^2+282r^3-
\end{displaymath}
\begin{displaymath}
-412r^4+333r^5-144r^6+28r^7)z^6+r(10-40r+56r^2-24r^3-14r^4+16r^5-4r^6)z^7
\Bigr].
\end{displaymath}
As mentioned above there exist several sources of relativistic corrections 
$|{\bf p}|^2/m_{1,2}^2$ in the expression (15).
The first part of terms appears from the expansion of the
gluon propagator and relativistic factors in the Dirac bispinors. The structure
of the spinor matrix element in this case is the same as in Eq.(17):
\begin{equation}
M_{21}(q,P,p',q')=\frac{2\alpha_s\sqrt{2M}}{3\sqrt{6}}\int\bar\Psi_{(\bar Q_1
Q_2),0}({\bf p}) \frac{d{\bf p}}{(2\pi)^3}\frac{(m_1+m_2)}{m_2(s-m_1^2)^2}
\times
\end{equation}
\begin{displaymath}
\times\Bigl\{\left[1-\frac{p^2}{k_0^2}+\frac{4(q p)^2}{k_0^4}-\frac{{\bf p}^2}
{8m_1^2}- \frac{{\bf p}^2}{8m_2^2}\right]\bar
u_2(p')2\hat{\tilde\epsilon}^\ast(v)(\hat q+m_1) \Gamma_\alpha v_1(q')+
\end{displaymath}
\begin{displaymath}
+\left[1-\frac{p^2}{k_0^2}+\frac{4(q p)^2}{k_0^4}-\frac{{\bf p}^2}{8m_1^2}-
\frac{{\bf p}^2}{8m_2^2}+\frac{2(q p)(p n)}{k_0^2(n q-\eta_1n P)}+
\frac{(p n)^2}{(n q-\eta_1n P)^2}\right]\times
\end{displaymath}
\begin{displaymath}
\times\frac{(s-m_1^2)}{(n q-\eta_1n P)}\bar u_2(p')\hat 
n\hat{\tilde\epsilon}^\ast
(v)(\hat v+1)\Gamma_\alpha v_1(q')\Bigr\}.
\end{displaymath}
Another part of corrections is determined both by the gluon propagator terms 
and
relativistic addenda $\hat{\tilde p}$ in the square brackets of Eq.(15):
\begin{equation}
M_{22}(q,P,p',q')=-\frac{2\alpha_s\sqrt{2M}}{3\sqrt{6}}\int\bar\Psi_{(\bar Q_1 
Q_2),0}
({\bf p})\frac{d{\bf p}}{(2\pi)^3}\frac{(m_1+m_2)}{m_2(s-m_1^2)^2}
\times
\end{equation}
\begin{displaymath}
\times\left[\frac{(n p )}{(n q-\eta_1n P)}+\frac{2q p}{k_0^2}\right]
\frac{(p_\sigma n_\lambda+p_\lambda n_\sigma)}{(n q-\eta_1 n P)}\bar
u_2(p')\gamma_\sigma\hat{\tilde\epsilon}^\ast(v)(\hat v+1)\gamma_\lambda(\hat 
q+m_1)
\Gamma_\alpha v_1(q'),
\end{displaymath}
\begin{equation}
M_{23}(q,P,p',q')=-\frac{2\alpha_s\sqrt{2M}}{3\sqrt{6}}\int\bar\Psi_{(\bar
Q_1 Q_2),0} ({\bf p})\frac{d{\bf
p}}{(2\pi)^3}\frac{(m_1+m_2)}{m_2(s-m_1^2)^2} \times
\end{equation}
\begin{displaymath}
\times\frac{(p_\sigma n_\lambda+p_\lambda n_\sigma)}{(n q-\eta_1 n P)}
\bar u_2(p')\gamma_\sigma\left[\frac{\hat{\tilde 
p}}{2m_2}\hat{\tilde\epsilon}^\ast
(v)(\hat v+1)+\hat{\tilde\epsilon}^\ast(v)(\hat v+1)\frac{\hat{\tilde p}}{2m_1}
\right]\gamma_\lambda(\hat q+m_1)\Gamma_\alpha v_1(q'),
\end{displaymath}
\begin{equation}
M_{24}(q,P,p',q')=-\frac{2\alpha_s\sqrt{2M}}{3\sqrt{6}}\int\bar\Psi_{(\bar
Q_1 Q_2),0} ({\bf p})\frac{d{\bf
p}}{(2\pi)^3}\frac{(m_1+m_2)}{m_2(s-m_1^2)^2} \times
\end{equation}
\begin{displaymath}
\times\left[g_{\lambda\sigma}-\frac{(k_\sigma n_\lambda+k_\lambda n_\sigma)}
{(nq-\eta_1 nP)}\right]\bar u_2(p')\gamma_\sigma\frac{\hat{\tilde p}}{2m_2}
\hat{\tilde\epsilon}^\ast(v)(\hat v+1)\frac{\hat{\tilde p}}{2m_1}\gamma_\lambda
(\hat q+m_1)\Gamma_\alpha v_1(q').
\end{displaymath}
Further transformations of the interference terms
$(M_{21}M^\ast_{2i}+ M_{2i}M^\ast_{21})$ ($i=2,3,4$) in the square modulus 
$|M|^2$ are performed by means of the system Reduce \cite{reduce}. The
relation
\begin{equation}
\int d\Omega_{{\bf p}}p_\mu p_\nu=(g_{\mu\nu}-v_\mu
v_\nu)\frac{{\bf P}^2- 3P^{0~2}}{9P^{0~2}}\int d\Omega_{{\bf
p}}{\bf p}^2,
\end{equation}
which accounts the orthogonality condition (3) is used for the integration
over the angle variables in the relative momentum space.
Omitting numerous intermediate expressions
appearing in the calculation $|M|^2$ which have sufficiently cumbersome
forms we present the final result for the relativistic correction in the
fragmentation function (16):
\begin{equation}
D^{rel}_{Q_1\to V(Q_1\bar 
Q_2)}(z)=\frac{8\alpha_s^2|\Psi(0)|^2}{27m_2^3}\frac{rz(1-z)^2}
{[1-(1-r)z]^6}v_2,
\end{equation}
\begin{equation}
v_2(z)=\frac{\left\langle{\bf 
p}^2\right\rangle}{36[1-(1-r)z]^2(1-r)^2m_2^2}\Bigl[-2+16r(2-3r)+
2(1-2r)(5-44r+30r^2)z+
\end{equation}
\begin{displaymath}
+(-23+18r-206r^2+508r^3-432r^4)z^2+8(1-r)(4+40r-11r^2-50r^3+66r^4)z^3-
\end{displaymath}
\begin{displaymath}
-2(1-r)^2(14+214r+47r^2-248r^3+120r^4)z^4+2(1-r)^3(7+99r+120r^2-230r^3+108r^4)
z^5-
\end{displaymath}
\begin{displaymath}
-(1-r)^4(3+26r+82r^2-84r^3+48r^4)z^6\Bigr],
\end{displaymath}
here $\left\langle{\bf p}^2\right\rangle$
is the special parameter determining the numerical value of 
relativistic effects (see the discussion before Eq.(60)).
Integrating expressions (19), (27) over $z$ we obtain the total fragmentation
probability:
\begin{equation}
\Omega_V=\int_0^1 D_{Q_1\to V(Q_1\bar 
Q_2)}(z)dz=\frac{8\alpha_s^2|\Psi(0)|^2}{405m_2^3(1-r)^6}
\left(f_{0v}(r)+\frac{W}{m_2}f_{1v}(r)+\frac{\left\langle{\bf 
p}^2\right\rangle}{m_2^2}f_{2v}(r)\right),
\end{equation}
\begin{equation}
f_{0v}(r)=24+85r-235r^2+300r^3-85r^4-89r^5+15r(7-4r+3r^2+10r^3+2r^4)\ln r,
\end{equation}
\begin{equation}
f_{1v}(r)=\frac{1}{42(1-r)^2}\Bigl[-2607+7185r+23576r^2-116018r^3+159670r^4-
170373r^5+153860r^6-
\end{equation}
\begin{displaymath}
-34906r^7-20387r^8
+210r(-42+223r-388r^2+236r^3-268r^4-43r^5+212r^6+28r^7)\ln
r\Bigr],
\end{displaymath}
\begin{equation}
f_{2v}(r)=\frac{1}{252(1-r)^2}\Bigl[488-3017r+48979r^2-201740r^3+136955r^4+
23597r^5-20958r^6+15696r^7-
\end{equation}
\begin{displaymath}
-105r(7-200r+347r^2+1194r^3-222r^4+156r^5+48r^6)\ln r\Bigr].
\end{displaymath}
The parameter $r=m_2/(m_1+m_2)$ is the ratio of the constituent mass of a
spectator quark to the mass of two quarks composing the heavy meson.
The contributions to the fragmentation functions (20)-(21) and (28)
give the distributions in the longitudinal momentum $z$ of the heavy
mesons. They are shown in Fig.2. We find it convenient to present here
the fragmentation functions of antiquarks for the comparison with experimental
data.
To extend the present calculations to the distributions in the transverse
momentum $p_T$ of the the heavy meson, the following relation between the
invariant mass $s$ of the fragmenting heavy quark, the
transverse momentum $p_T$ and longitudinal momentum $z$ is taken into account
\cite{Close}:
\begin{equation}
s(z,t)=\frac{M^2+p_T^2}{z}+\frac{m_2^2+p_T^2}{1-z}.
\end{equation}
Introducing further the dimensionless variable $t=p_T/(m_1+m_2)$, we can 
determine
the $p_T$-dependent relativistic corrections to the fragmentation
probability $D_{Q_1\to V(Q_1\bar Q_2)}(t)$ \cite{B4}:
\begin{equation}
D_{Q_1\to V(Q_1\bar
Q_2)}(t)=\int_0^1dz\frac{2M^2t}{z(1-z)}D_{Q_1\to V(Q_1\bar
Q_2)}\left(z,s(z,t)\right)=
\end{equation}
\begin{displaymath}
=\frac{4\alpha_s^2|\Psi(0)|^2r}{27m_2^3(1-r)^6t^6}
\left[D_{0v}(t)+\frac{W}{m_2}D_{1v}(t)+\frac{\left\langle{\bf 
p}^2\right\rangle}{m_2^2}D_{2v}(t)\right],
\end{displaymath}
\begin{equation}
D_{0v}(t)=\Bigl\{ (-30r^3 +30r^4)t +(61r - 33r^2 - 48r^3 +20r^4)t^3
+(-5+ 13r - 16r^2 + 4r^3 +4r^4)t^5 +
\end{equation}
\begin{displaymath}
+\arctan\frac{t(1-r)}{r+t^2}\Bigl[30r^4 +(-99r^2 -66r^3 +30r^4)t^2
+(9 +20r +r^2 +22r^3 +8r^4)t^4 +
\end{displaymath}
\begin{displaymath}
+(9 -12r + 4r^2 + 8r^3)t^6\Bigr] +\ln(r)\Bigl[-96r^3t +(48r +
56r^2 +16r^3)t^3\Bigr]+
\end{displaymath}
\begin{displaymath}
+\ln\frac{(1+t^2)}{(r^2+t^2)}\Bigl[-48r^3t +(24r + 28r^2 +8r^3)t^3 +
(-4+ 8r -4r^2)t^7\Bigr]\Biggr\},
\end{displaymath}
\begin{equation}
D_{1v}(t)=\frac{1}{3t^2(1-r)^2}
\Bigl\{-\frac{24rt(1-r)^6}{1+t^2}+(24r-144r^2+360r^3 - 480r^4 + 45r^5 +
486r^6-291r^7)t+
\end{equation}
\begin{displaymath}
+(-24r +144r^2 -633r^3+2115r^4-1233r^5-1266r^6+897r^7)t^3+
\end{displaymath}
\begin{displaymath}
+(-18r+1130r^2 - 4129r^3 + 3938r^4 + 595r^5 -1994r^6 + 478r^7)t^5+
\end{displaymath}
\begin{displaymath}
+(30-215r +581r^2 - 742r^3+738r^4-596r^5+148r^6+56r^7)t^7+
\end{displaymath}
\begin{displaymath}
+(60r -255r^2+360r^3-195r^4+30r^5)t^9\Bigr]+
\end{displaymath}
\begin{displaymath}
+\arctan\frac{t(1-r)}{(r+t^2)}\Bigl[315r^6-315r^7+(90r^4-1650r^5-345r^6+
960r^7)t^2+
\end{displaymath}
\begin{displaymath}
+(-63r^2 -1800r^3 +5646r^4 -714r^5 -3495r^6 +741r^7)t^4 +
\end{displaymath}
\begin{displaymath}
+(-54 +237r - 33r^2 - 807r^3 + 642r^4 -621r^5 +423r^6+150r^7)t^6 +
\end{displaymath}
\begin{displaymath}
+(-54 + 297r -507r^2 + 216r^3 - 252r^4 +147r^5 +126r^6)t^8+
\end{displaymath}
\begin{displaymath}
+(-60r +195r^2 -165r^3 +30r^4)t^{10}\Bigr]+
\end{displaymath}
\begin{displaymath}
+\ln(r)\Bigl[(-576r^5 +288r^7)t +(-432r^3 +48r^4 + 5856r^5 -
4416r^6 -48r^7)t^3 +
\end{displaymath}
\begin{displaymath}
+(-144r +1296r^2 -2136r^3 -768r^4 +1320r^5 +432r^6)t^5\Bigr]+
\end{displaymath}
\begin{displaymath}
+\ln\frac{(1+t^2)}{(r^2+t^2)}\Bigr[(-288r^5+144r^7)t+(-216r^3+24r^4+2928r^5
-2208r^6 -24r^7)t^3+
\end{displaymath}
\begin{displaymath}
+(-72r +648r^2-1068r^3-384r^4+660r^5+216r^6)t^5 +
\end{displaymath}
\begin{displaymath}
+(24-156r+312r^2-192r^3+72r^4-60r^5)t^9\Bigr]\Bigr\},
\end{displaymath}
\begin{equation}
D_{2v}(t)=\frac{1}{36t^2(1-r)^2}\Bigl\{-420r^5(1-r)t+
(6074r^3-20646r^4+11976r^5+2596r^6)t^3+
\end{equation}
\begin{displaymath}
+(-573r+2309r^2+5008r^3-5180r^4-1636r^5+72r^6)t^5+
\end{displaymath}
\begin{displaymath}
+(5-189r+1124r^2-960r^3+68r^4+48r^5-96r^6)t^7+
\end{displaymath}
\begin{displaymath}
+\arctan\frac{t(1-r)}{r+t^2}\Bigl[420r^6+(-7470r^4+15480r^5+5040r^6)t^2+
\end{displaymath}
\begin{displaymath}
+(1887r^2-1758r^3-20334r^4-6936r^5+900r^6)t^4+
\end{displaymath}
\begin{displaymath}
+(-9+100r+1303r^2-1614r^3-632r^4+216r^5-168r^6)t^6+
\end{displaymath}
\begin{displaymath}
+(-9+108r-1336r^2-96r^3+64r^4-192r^5)t^8\Bigr]+
\end{displaymath}
\begin{displaymath}
+\ln(r)\Bigl[(-3072r^5+4608r^6)t+(6240r^3-10688r^4-24256r^5+768r^6)t^3+
\end{displaymath}
\begin{displaymath}
+(-240r+712r^2+5232r^3+3392r^4+384r^5)t^5\Bigr]+
\end{displaymath}
\begin{displaymath}
+\ln\frac{(r^2+t^2)}{(t^2+1)}\Bigl[(-1536r^5+2304r^6)t+
(3120r^3-5344r^4-12128r^5+384r^6)t^3+
\end{displaymath}
\begin{displaymath}
+(-120r+356r^2+2616r^3+1696r^4+192r^5)t^5+
\end{displaymath}
\begin{displaymath}
+(4+72r+324r^2-208r^3+96r^4)t^9
\Bigr]\Bigr\}.
\end{displaymath}
The leading order contribution (35) coincides with the result of
Ref.\cite{B4} and the two other terms (36) and (37) determine the
relativistic and bound state corrections. The functions (35)-(37) are plotted 
in Fig.4.

Let us next consider the calculation of the fragmentation functions into 
pseudoscalar
heavy mesons $\eta_c$, $B_c$, $\eta_b$. The general expression of the
fragmentation amplitude consists of three terms:
\begin{equation}
M_3=\frac{2\sqrt{2M}\alpha_s|\Psi(0)|}{3\sqrt{6}}\frac{(m_1+m_2)}{m_2
(s-m_1^2)^2}\Bigl[a_1\bar u_2(p')\gamma_5(\hat q+m_1)\Gamma_\alpha v_1(q')+
\end{equation}
\begin{displaymath}
+a_2\bar u_2(p')\gamma_5\Gamma_\alpha v_1(q')+a_3\bar u_2(p')\gamma_5
\hat n(\hat q+m_1)\Gamma_\alpha v_1(q')\Bigr],
\end{displaymath}
(the fourth possible term $\bar u_2(p')\gamma_5\hat n\Gamma_\alpha v_1(q')$
does not contribute to the fragmentation function)
where the coefficients $a_i$ (i=1,2,3) contain the leading order contribution
and corrections proportional to $\left\langle{\bf p}^2\right\rangle$ and $W$:
\begin{equation}
a_1=1-\frac{2}{9}\frac{\left\langle{\bf 
p}^2\right\rangle}{(s-m_1^2)r}+\frac{\left\langle{\bf 
p}^2\right\rangle}{m_1^2}
\Bigl[\frac{5}{24}+\frac{2}{9}r-\frac{5}{9}r^2+
\end{equation}
\begin{displaymath}
+\frac{1}{1-z(1-r)}\left(\frac{1}{2}zr^3-\frac{2}{3}zr^2-\frac{1}{6}r^2-
\frac{1}{18}rz+\frac{7}{18}r+\frac{2}{9}z-\frac{2}{9}\right)\Bigr]+
\end{displaymath}
\begin{displaymath}
+\frac{2(m_1+m_2)W}{(s-m_1^2)}(1-r)+\frac{W}{(m_1+m_2)}\left(2-\frac{1}{r}-
\frac{z}{1-z(1-r)}+\frac{2rz}{1-z(1-r)}\right),
\end{displaymath}
\begin{equation}
a_2=1+\frac{\left\langle{\bf 
p}^2\right\rangle}{m_1^2}\left(-\frac{1}{6}r^2+\frac{5}{18}r-\frac{17}{12}\right)+
\frac{2}{9}\frac{\left\langle{\bf 
p}^2\right\rangle}{m_2^2}\frac{r^2z^2}{[1-z(1-r)]^2}+
\end{equation}
\begin{displaymath}
+\frac{\left\langle{\bf 
p}^2\right\rangle}{m_1^2}\frac{1}{1-z(1-r)}\left(-\frac{2}{9}r^2z+\frac{2}{9}
r^2+\frac{1}{3}rz-\frac{1}{3}r-\frac{1}{9}z+\frac{1}{9}\right)+
\end{displaymath}
\begin{displaymath}
+\frac{\left\langle{\bf 
p}^2\right\rangle}{m_1^2}\frac{m_2^2}{(s-m_1^2)}\left(\frac{5}{9}r-
\frac{5}{9rz}+\frac{1}{r}+\frac{2}{9r^2z}-\frac{2}{9r^2}+\frac{1}{3z}-
\frac{4}{3}\right)-\frac{4}{9r}\frac{\left\langle{\bf 
p}^2\right\rangle}{(s-m_1^2)}+
\end{displaymath}
\begin{displaymath}
+\frac{2(m_1+m_2)W}{(s-m_1^2)}(1-r)+\frac{W}{(m_1+m_2)}\left(2-\frac{1}{r}\right),
\end{displaymath}
\begin{equation}
a_3=1-\frac{\left\langle{\bf 
p}^2\right\rangle}{9m_1^2}\left(r+\frac{1}{8}\right)+
\frac{\left\langle{\bf 
p}^2\right\rangle}{m_1m_2}\frac{rz}{1-z(1-r)}+\frac{2r^2z^2}{9}\frac{\left\langle{\bf 
p}^2\right\rangle}{m_2^2}-
\end{equation}
\begin{displaymath}
-\frac{2}{9r}\frac{\left\langle{\bf p}^2\right\rangle}{(s-m_1^2)}+
\frac{2(m_1+m_2)W}{(s-m_1^2)}(1-r)+\frac{W}{(m_1+m_2)}\left(2-
\frac{1}{r}\right).
\end{displaymath}
The exact expressions (39)-(41) are obtained on the basis of Eq.(15)
($\tilde \epsilon\to\gamma_5$) keeping the terms of the second
order over relative momentum $p$ and binding energy corrections.
The squared modulus amplitude $|M_3|^2$ leads by using the definition (16) 
after the integration over $s$
to the following fragmentation distribution for the production of
the pseudoscalar heavy mesons:
\begin{equation}
D_{Q_1\to P(Q_1\bar
Q_2)}(z)=\frac{8\alpha_s^2|\Psi(0)|^2}{81m_2^3}
\frac{rz(1-z)^2}{[1-(1-r)z]^6}(p_0+p_1+p_2),
\end{equation}
\begin{equation}
p_0=6+18(2r-1)z+(21-74r+68r^2)z^2-2(1-r)(6-19r+18r^2)z^3+
3(1-r)^2(1-2r+2r^2)z^4,
\end{equation}
\begin{equation}
p_1=\frac{W}{m_2[1-(1-r)z]^2}\Bigl\{-12+6r+60z+24r(-5+2r)z+[-126+r(425-482r+
218r^2)]z^2+
\end{equation}
\begin{displaymath}
+2[72+r(-329+2r(298+r(-283+131r)))]z^3-2(1-r)[48+r(-219+r(410+7r(-61+
33r))]z^4
\end{displaymath}
\begin{displaymath}
+2(1-r)^2[18+r(-79+2r(69+13r(-5+3r)))]z^5+(r-1)^3[6+r(-25+r(35+6r(-4+3r)))
]z^6\Bigr\},
\end{displaymath}
\begin{equation}
p_2=\frac{\left\langle{\bf 
p}^2\right\rangle}{36(1-r)^2m_2^2[1-(1-r)z]^2}\Bigl\{
-6-48r-6[-5+2r(-3+r(-3+22r))]z-
\end{equation}
\begin{displaymath}
-[63+2r(-191+r(431+4r(-197+146r)))]z^2+8(1-r)[9+r(-108+r(217+5r(-42+29r)))]z^3-
\end{displaymath}
\begin{displaymath}
-2(1-r)^2[24+r(-390+r(697+4r(-53+5r)))]z^4+2(r-1)^3[-9+r(155+2r\times
\end{displaymath}
\begin{displaymath}
(-122+r(-7+36r)))]z^5+(1-r)^4(-3+42r-58r^2+24r^4)z^6\Bigr\}.
\end{displaymath}
Integrating expressions (43)-(45) over $z$ we obtain the total
fragmentation probability for the pseudoscalar mesons:
\begin{equation}
\Omega_P=\int_0^1 D_{Q_1\to P(Q_1\bar 
Q_2)}(z)dz=\frac{8\alpha_s^2|\Psi(0)|^2}{405m_2^3(1-r)^6}
\left(f_{0p}(r)+\frac{W}{m_2}f_{1p}(r)+\frac{\left\langle{\bf 
p}^2\right\rangle}{m_2^2}f_{2p}(r)\right),
\end{equation}
\begin{equation}
f_{0p}(r)=8+5r+215r^2-440r^3+265r^4-53r^5+15r(1+8r+r^2-6r^3+2r^4)\ln r,
\end{equation}
\begin{equation}
f_{1p}(r)=\frac{1}{42(1-r)}\Bigl[897-3640r+8400r^2-59850r^3+147105r^4-
132762r^5+46830r^6-
\end{equation}
\begin{displaymath}
-6980r^7+210r(6-3r-129r^2+115r^3+123r^4-84r^5+18r^6)\ln r\Bigr],
\end{displaymath}
\begin{equation}
f_{2p}(r)=\frac{1}{756(1-r)^2}\Bigl[614-9051r-9681r^2+106470r^3-73815r^4-
45129r^5+
\end{equation}
\begin{displaymath}
+41146r^6-10554r^7+105r(-3-204r+281r^2+798r^3-402r^4+48r^5+24r^6)\ln r\Bigr].
\end{displaymath}
The transverse momentum fragmentation functions for the production of
pseudoscalar mesons $\eta_c$, $B_c$ and $\eta_b$ can be derived in a similar
way as for the vector mesons. The corresponding expressions are given as 
follows:
\begin{equation}
D_{Q_1\to P(Q_1\bar
Q_2)}(t)=\int_0^1dz\frac{2M^2t}{z(1-z)}D_{Q_1\to P(Q_1\bar Q_2)}
\left(z,s(z,t)\right)=
\end{equation}
\begin{displaymath}
=\frac{4\alpha_s^2|\Psi(0)|^2r}{81m_2^3(1-r)^6t^6}
\left[D_{0p}(t)+\frac{W}{m_2}D_{1p}(t)+\frac{\left\langle{\bf 
p}^2\right\rangle}{m_2^2}D_{2p}(t)\right],
\end{displaymath}
\begin{equation}
D_{0p}(t)=\Bigl\{-(1-r)t[30r^3-r(61-20r+28r^2)t^2-(3-48r+48r^2-12r^3)t^4]+
\end{equation}
\begin{displaymath}
+3\arctan\frac{t(1-r)}{r+t^2}\Bigl[10r^4-3r^2(11+2r+2r^2)t^2+(3+4r+19r^2-
6r^3)t^4+
\end{displaymath}
\begin{displaymath}
+(3+12r-20r^2+8r^3)t^6)\Bigr] -24rt\ln(r)[4r^2-(2+r+2r^2)t^2]-
\end{displaymath}
\begin{displaymath}
-12t\ln\frac{(1+t^2)}{(r^2+t^2)}[4r^3-r(2+r+2r^2)t^2+(1-r)^2t^6]\Bigr\},
\end{displaymath}
\begin{equation}
D_{1p}(t)=\frac{1}{t^2(1-r)}
\Bigl\{\frac{24rt(1-r)^4(2r-1)}{1+t^2}+3r(8-48r+112r^2-128r^3+37r^4+19r^5)t+
\end{equation}
\begin{displaymath}
+r(-24+144r-691r^2+1032r^3+84r^4-545r^5)t^3+
\end{displaymath}
\begin{displaymath}
+r(10+360r-1180r^2+1037r^3-495r^4+268r^5)t^5-
\end{displaymath}
\begin{displaymath}
-3(2-19r+117r^2-300r^3+320r^4-140r^5+20r^6)t^7+
\end{displaymath}
\begin{displaymath}
+3\arctan\frac{t(1-r)}{(r+t^2)}\Bigl[35r^6-10r^4(-13+15r+25r^2)t^2+
\end{displaymath}
\begin{displaymath}
+r^2(-15-281r+501r^2+96r^3+79r^4)t^4+
\end{displaymath}
\begin{displaymath}
+(-6+31r-7r^2+79r^3-245r^4+42r^5+8r^6)t^6-
\end{displaymath}
\begin{displaymath}
-(6-11r-68r^2+208r^3-160r^4+40r^5)t^8+
\end{displaymath}
\begin{displaymath}
+24rt\ln(r)[-16r^5+2r^2(-7-10r+38r^2+4r^3)t^2+(-2+20r-21r^2-5r^3-
22r^4)t^4]+
\end{displaymath}
\begin{displaymath}
+12t\ln\frac{(r^2+t^2)}{1+t^2}[16r^6-2r^3(-7-10r+38r^2+4r^3)t^2+
\end{displaymath}
\begin{displaymath}
+r(2-20r+21r^2+5r^3+22r^4)t^4-(1-r)^2(2-5r+5r^2)t^8]\Bigr\},
\end{displaymath}
\begin{equation}
D_{2p}(t)=\frac{1}{36t^2(1-r)^2}\Bigl\{-420r^5(1-r)t+
2r^3(1957-7323r+4368r^2+998r^3)t^3-
\end{equation}
\begin{displaymath}
-r(429-2629r+152r^2-396r^3+1684r^4+760r^5)t^5+
\end{displaymath}
\begin{displaymath}
+3(-1+65r-308r^2+360r^3-108r^4-24r^5+16r^6)t^7+
\end{displaymath}
\begin{displaymath}
+\arctan\frac{t(1-r)}{r+t^2}\Bigl[(420r^6+(-5310r^4+11640r^5+4440r^6)t^2+
\end{displaymath}
\begin{displaymath}
+(1023r^2-4158r^3-9486r^4-4080r^5-1764r^6)t^4+
\end{displaymath}
\begin{displaymath}
+(-9-324r-9r^2+1674r^3+1632r^4-360r^5+72r^6)t^6+
\end{displaymath}
\begin{displaymath}
+(-9-36r+744r^2-552r^3-24r^4+96r^5)t^8\Bigr]+
\end{displaymath}
\begin{displaymath}
+24rt\ln(r)\Bigl[-128r^4+192r^5-4r^2(-35+110r+110r^2+26r^3)t^2+
\end{displaymath}
\begin{displaymath}
+(-10-r+150r^2+152r^3+100r^4+4r^5)t^4\Bigr]+
\end{displaymath}
\begin{displaymath}
+12t\ln\frac{(1+t^2)}{(r^2+t^2)}\Bigl[-128r^5+192r^6-4r^3(-35+110r+110r^2+
26r^3)t^2+
\end{displaymath}
\begin{displaymath}
+r(-10-r+150r^2+152r^3+100r^4+4r^5)t^4-(-1+14r-13r^2-4r^3+r^4)t^8)\Bigr]\Bigr\}.
\end{displaymath}
The leading order function (51) coincides with the result of
Ref.\cite{B4}, and the two other functions (52) and (53) determine the
relativistic and binding energy corrections. Functions (51)-(53)
are plotted in Fig.5. The obtained results for the transverse momentum 
distributions are valid for the transverse momentum $p_T$ up to values of 
order of the 
meson mass. The further growth of the momentum $p_T$ demands the consideration 
of the omitted corrections of order $O(p_T/M_Z)$.

\section{Quasipotential quark model}

To estimate numerical values of the investigated effects in the heavy quark 
fragmentation we used the relativistic quark model.
In the quasipotential approach the bound state of a quark and
antiquark is described by the Schr\"odinger type equation
\cite{rqm4}
\begin{equation}
\label{quas}
{\left(\frac{b^2(M)}{2\mu_{R}}-\frac{{\bf p}^2}{2\mu_{R}}\right)
\Psi_0({\bf p})} =\int\frac{d{\bf q}}{(2\pi)^3}V({\bf p,q},M)
\Psi_0({\bf q}),
\end{equation}
where the relativistic reduced mass is
\begin{equation}
\mu_{R}=\frac{E_1E_2}{E_1+E_2}=\frac{M^4-(m^2_1-m^2_2)^2}{4M^3},
\end{equation}
and the particle energies $E_1$, $E_2$ are given by
\begin{equation}
\label{ee}
E_1=\frac{M^2-m_2^2+m_1^2}{2M}, \quad E_2=\frac{M^2-m_1^2+m_2^2}{2M},
\end{equation}
here $M=E_1+E_2$ is the bound state mass,
$m_{1,2}$ are the masses of heavy quarks ($Q_1$ and $Q_2$) which form
the meson, and ${\bf p}$  is their relative momentum.
In the center of mass system the relative momentum squared on mass shell
reads
\begin{equation}
{b^2(M)}=\frac{[M^2-(m_1+m_2)^2][M^2-(m_1-m_2)^2]}{4M^2}.
\end{equation}

\begin{table}
\caption{\label{t1} Basic parameters of the relativistic quark model
and the total fragmentation probabilities $\Omega$ for the reactions $\bar 
c\to \eta_c,J/\Psi$, $\bar b\to B_c,B_c^\ast$, $\bar b\to\eta_b,\Upsilon$.}
\bigskip
\begin{ruledtabular}
\begin{tabular}{|c|c|c|c|c|c|c|c|}   \hline
State  & Particle & Mass,~$GeV$  & $\bar\alpha_V$ & Bound state &
$\Psi(0),~GeV^{3/2}$ &$\left\langle{\bf p}^2\right\rangle,~GeV^2$ & $\Omega$ \\
$n^{2S+1}L_J$&  &\cite{PDG}  &  & energy,~GeV &  &  & \\ \hline
$1^1S_0$ & $\eta_c$ & 2.980 & 0.451 &-0.120  &0.27   & 0.5 & $1.49\times
10^{-4}$ \\  \hline
$1^3S_1$ & $J/\Psi$ & 3.097 & 0.451 &-0.003  &0.27   &0.5  & $1.28\times
10^{-4}$ \\  \hline
$1^1S_0$ & $B_c$ & 6.270$^a$ & 0.361 &-0.160  &0.33   &0.7 & $3.49\times 
10^{-4}$  \\  \hline
$1^3S_1$ & $B_c^\ast$ & 6.332$^a$ & 0.361 &-0.098  &0.33   &0.7 & $4.89\times
10^{-4}$  \\  \hline
$1^1S_0$ & $\eta_b$ & 9.400\footnote{This value was obtained in the 
Ref.\cite{rqm1}} & 0.267 &-0.360  &0.46   & 1.4 & $0.14\times 10^{-4}$ \\
\hline
$1^3S_1$ & $\Upsilon$ & 9.460 & 0.267 &-0.300  &0.46   &1.4 & $0.15\times 
10^{-4}$  \\  \hline
\end{tabular}
\end{ruledtabular}
\end{table}

\begin{figure}[htbp]
\vspace{-6mm}
\begin{center}
$\bar c\to J/\Psi$
\end{center}
\centering
\includegraphics{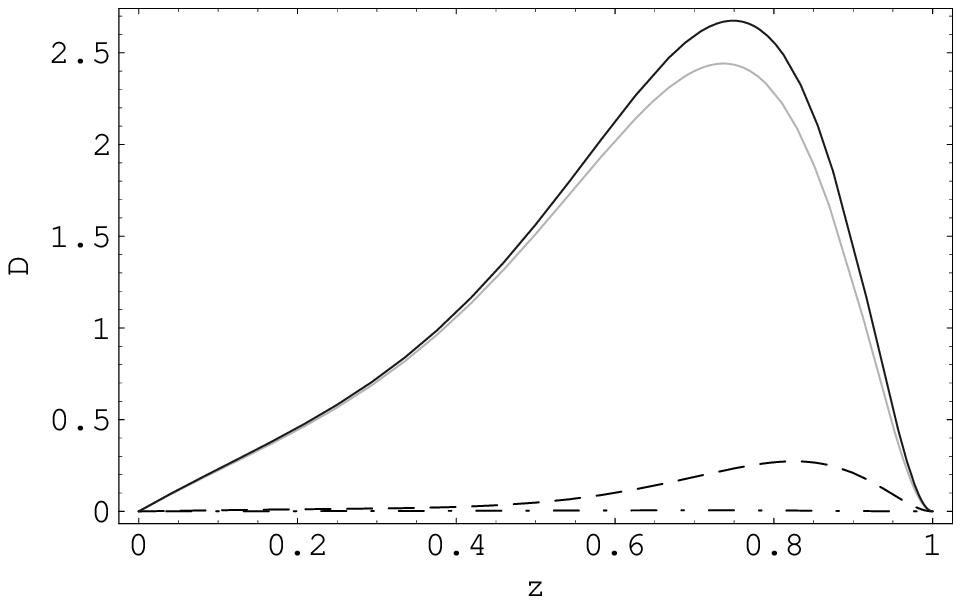}
\vspace{-2mm}
\begin{center}
$\bar b\to B_c^\ast$
\end{center}
\includegraphics{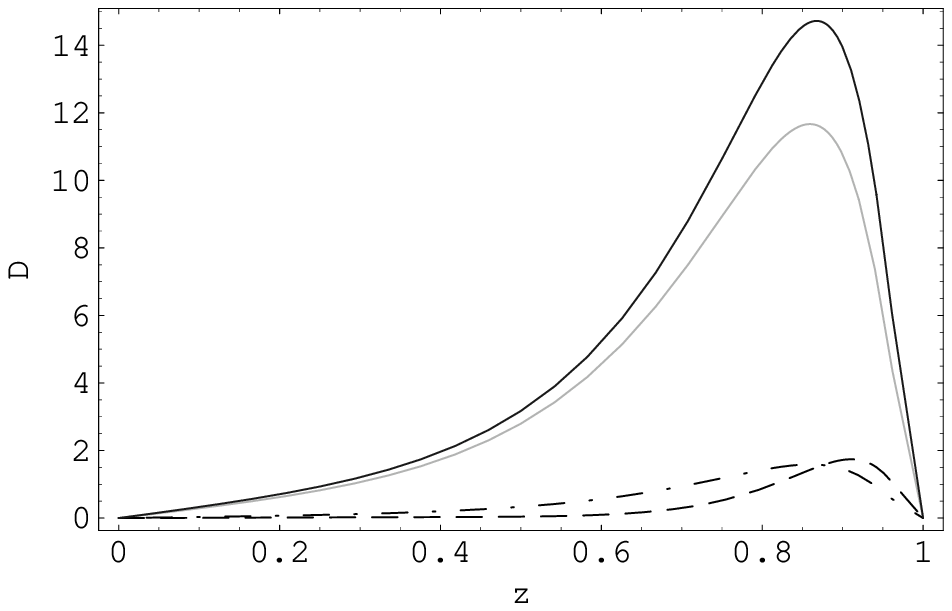}
\vspace{-2mm}
\begin{center}
$\bar b\to\Upsilon$
\end{center}
\includegraphics{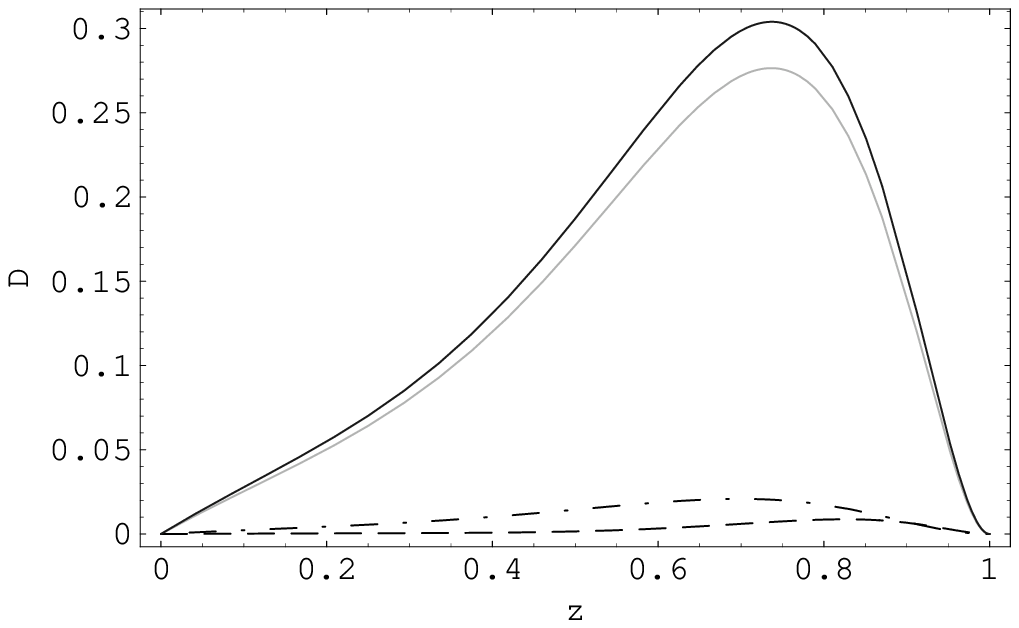}
\vspace{-4mm}
\caption{The contributions to the fragmentation functions
$D(\bar c\to J/\Psi)(z)$,
$D(\bar b\to B_c^\ast)(z)$,
$D(\bar b\to \Upsilon)(z)$.
The thick solid line shows the total fragmentation
function, the dashed line shows the relativistic correction (27), the
dotdashed line shows the correction proportional to the binding
energy $W$ (21). 
The thin solid
line corresponds to the distributions without corrections.
All functions have been multiplied by a factor $10^{4}$.}
\end{figure}

\begin{figure}[htbp]
\vspace{-6mm}
\begin{center}
$\bar c\to \eta_c$
\end{center}
\centering
\includegraphics{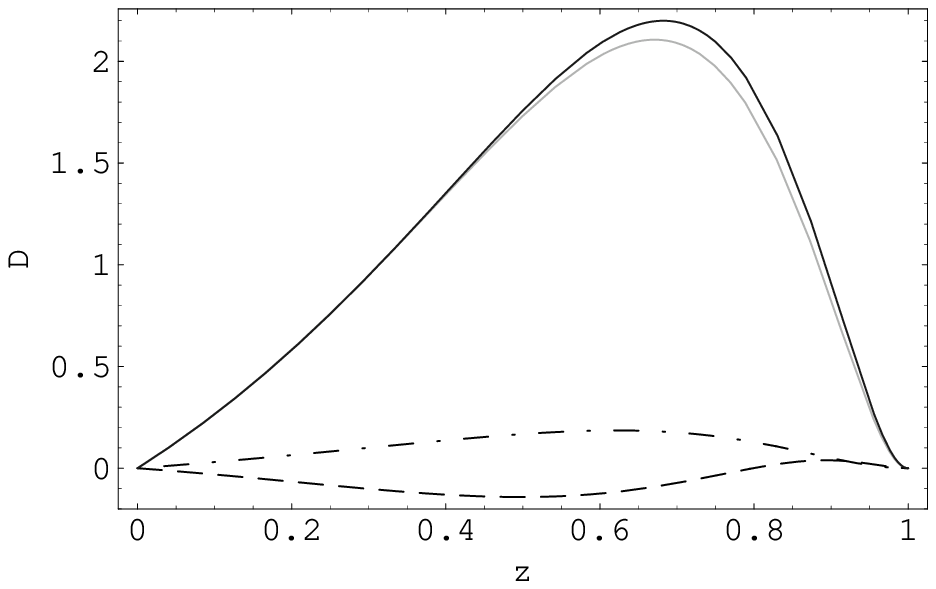}
\vspace{-2mm}
\begin{center}
$\bar b\to B_c$
\end{center}
\includegraphics{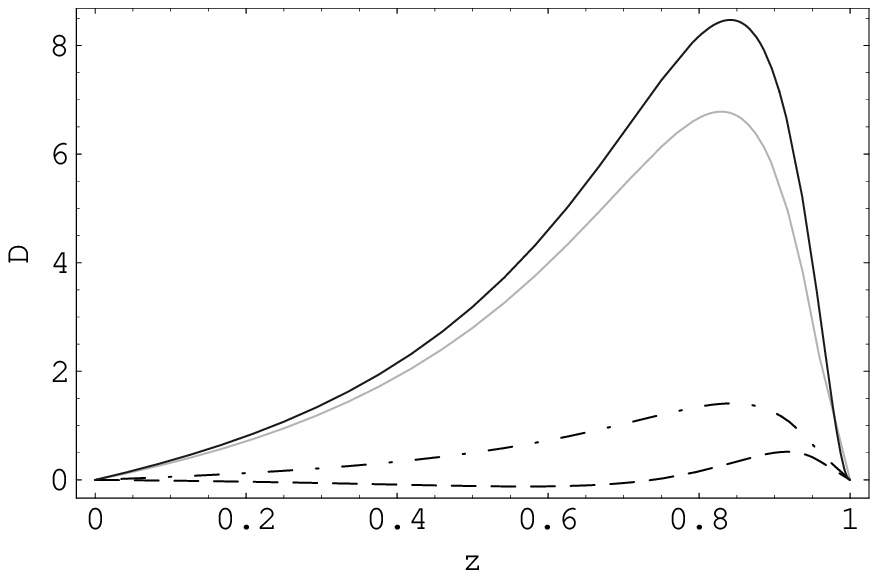}
\vspace{-2mm}
\begin{center}
$\bar b\to\eta_b$
\end{center}
\includegraphics{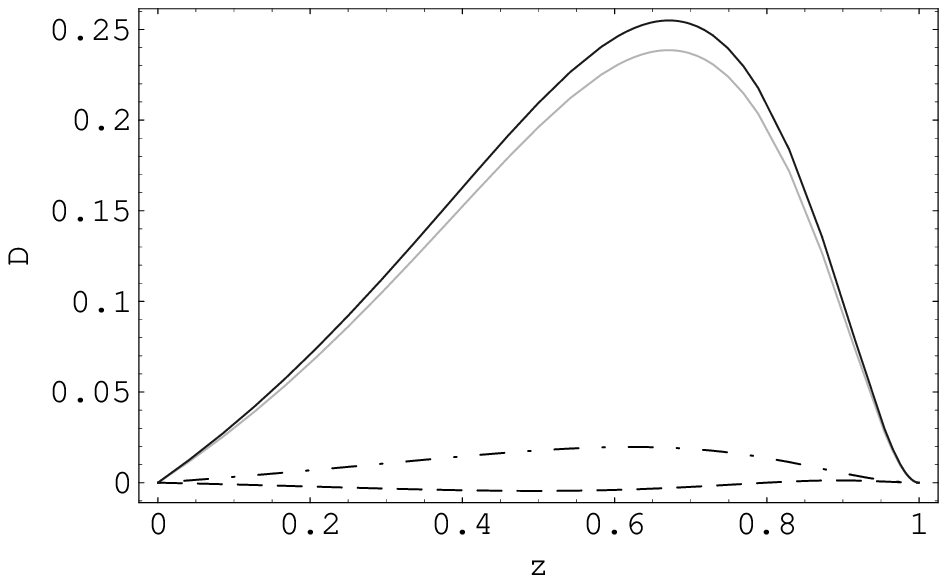}
\vspace{-4mm}
\caption{The contributions to the fragmentation functions $D(\bar
c\to \eta_c)(z)$, $D(\bar b\to B_c)(z)$, $D(\bar b\to \eta_b)(z)$.
The thick solid line shows the total fragmentation function, the dashed
line shows the relativistic correction (45), the dotdashed line shows
the correction proportional to the binding energy $W$ (44). 
The thin solid
line corresponds to the distributions without corrections.
All functions have been multiplied by a factor $10^{4}$.}
\end{figure}

The kernel $V({\bf p,q},M)$ in Eq.~(\ref{quas}) is the quasipotential
operator of the quark-antiquark interaction. Within an effective field theory 
(NRQCD) the quark-antiquark potential was constructed in Ref.\cite{NR1,NR2}
by the perturbation theory improved by the renormalization group resummation 
of large logarithms. In the quasipotential quark model the kernal $V({\bf 
p},{\bf 
q},M)$ is constructed phenomenologically with the help of the off-mass-shell 
scattering amplitude, projected onto the positive
energy states. The heavy quark-antiquark potential with the account of
retardation effects and the one loop radiative corrections can be presented
in the form of a sum of spin-independent and spin-dependent parts.
Explicit expressions for it are given in Refs. \cite{rqm2,rqm3}.
Taking into account the accuracy of the calculation of relativistic
corrections to the fragmentation probabilities, we can use for the description
of the bound system $(Q_1 \bar Q_2)$ the following simplified interaction 
operator in the
coordinate representation:
\begin{equation}
\tilde V(r)=-\frac{4}{3}\frac{\bar\alpha_V(\mu^2)}{r}+Ar+B,
\end{equation}
where the parameters of the linear potential $A=0.18~GeV^2$, $B=-0.16~GeV$,
\begin{equation}
\bar\alpha_V(\mu^2)=\alpha_s(\mu^2)\left[1+\left(\frac{a_1}{4}+\frac{\gamma_E
\beta_0}{2}\right)\frac{\alpha_s(\mu^2)}{\pi}\right],
\end{equation}
\begin{displaymath}
a_1=\frac{31}{3}-\frac{10}{9}n_f,~~ \beta_0=11-\frac{2}{3}n_f.
\end{displaymath}
Here $n_f=3$ is the number of flavors and $\mu$ is a
renormalization scale. All the parameters of the model like quark
masses, parameters of the linear confining potential $A$ and $B$,
mixing coefficient $\varepsilon$ and anomalous chromomagnetic
quark moment $\kappa$ entering in the quasipotential $V({\bf p},{\bf q},M)$
were fixed from the analysis of heavy
quarkonium masses \cite{rqm1,rqm2,rqm3} and radiative decays
\cite{rqm2}. The heavy quark masses $m_b=4.88$ GeV,
$m_c=1.55$ GeV and the parameters of the linear potential
$A=0.18$ GeV$^2$ and $B=-0.16$ GeV have standard values of the
quark models. Solving the Schr\"odinger-like quasipotential
equation we obtain an initial expression for the bound state wave
functions in the case of $(c\bar c)$, $(\bar b c)$ and $(\bar b b)$ systems.
For numerical estimations of relativistic effects in the production
of heavy mesons via heavy quark fragmentation we need the values
of the wave functions at the origin, the bound state energy and
the parameter of relativistic effects: $\int{\bf p}^2\bar{\Psi}_0({\bf
p})d{\bf p}/(2\pi)^3$. Note that this integral would diverge in the high 
momentum
region. Obviously, the reason of this divergence is
connected with the used expansion of the integral function in the
basic equation (13) over the ratio ${\bf p}^2/m^2$. In the
coordinate representation this divergence would appear when we set
${\bf r}=0$ in the Coulomb part of the potential (58). Different
regularization prescriptions are commonly used in this case
\cite{aKM,Labelle,KM,Khan}. An approach to the calculation of this integral 
was formulated in Ref.\cite{KM1} for solving the orthopositronium decay 
problem in quantum electrodynamics. 
Their prescription is in an agreement with the
calculations carried out in the effective field theories \cite{Adkins}.
Unfortunately, in the investigation of the bound states of heavy quarks we 
cannot use it because the valid wave function asymptotics at $p\to \infty$ is 
not the 
Coulomb-like. So, to fix the value of the relativistic correction we explore 
the dimensional regularization scheme where the scaleless momentum integral
$\int V({\bf p}-{\bf q})\Psi({\bf q})\frac{d^d{\bf q}}{(2\pi)^d}\frac{d^d{\bf 
p}}{(2\pi)^d}$ related to the problem vanishes \cite{aKM,Labelle,CMY}. Then we 
can express 
the necessary quantity in the form:
\begin{equation}
\left\langle {\bf p}^2\right\rangle\equiv \frac{1}{\Psi(0)}\int \frac{d^d{\bf 
p}}{(2\pi)^d}{\bf p}^2\bar\Psi_0({\bf p})=2\mu_R \tilde W+2\mu_R |B|.
\end{equation}
The solutions of the Schr\"odinger-like equation (54) with the potential (58) 
determine the energy spectrum $\tilde W$ of the heavy quark system and lead to 
the numerical values of the parameter (60) for the bound states $(\bar cc)$,
$(\bar bb)$
and $(\bar b c)$ which are presented in Table I. They are in qualitative
agreement with the other possible approach for the estimation of the value (60)
based on the natural regularization directly connected with the relativistic 
structure factors entering in the fragmentation amplitude (13). 
Heavy quark symmetry predicts that the wave functions of the vector and 
pseudoscalar states are different due to corrections of order $v_Q^2$. The
analogous statement is valid for the parameter $\left\langle {\bf 
p}^2\right\rangle$. Nevertheless, in this study we neglect this difference
and write in Table I equal values for $\Psi(0)$ and $\left\langle {\bf 
p}^2\right\rangle$ for $V$- and $P$-mesons. Our value
$\left\langle {\bf p}^2\right\rangle=0.5~GeV^2$ for $(c\bar c)$-states is
slightly smaller than $\left\langle {\bf p}^2\right\rangle=0.7~GeV^2$ used
in Ref.\cite{Bashir} where it was fixed from the analysis of the quarkonium
decay rates. 
The theoretical uncertainty of the obtained values
$\left\langle {\bf p}^2\right\rangle$ in the Table I is determined by
perturbative and nonperturbative corrections to the quasipotential
\cite{rqm1,rqm2} and is not exceeding $30\%$.

\begin{figure}[htbp]
\vspace{-6mm}
\begin{center}
$\bar c\to J/\Psi$
\end{center}
\centering
\includegraphics{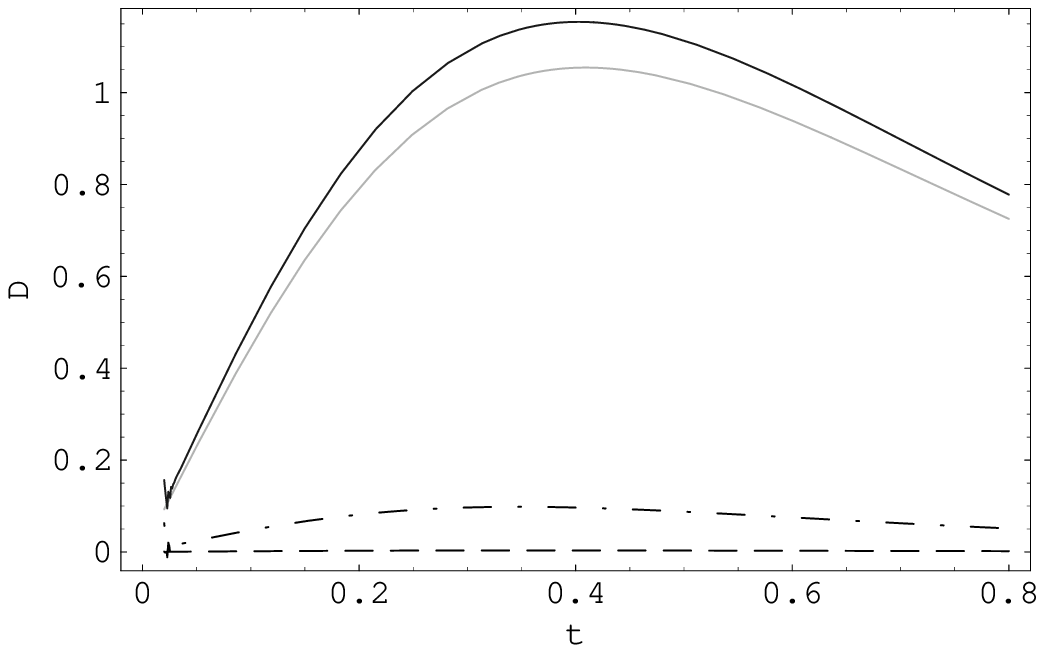}
\vspace{-2mm}
\begin{center}
$\bar b\to B_c^\ast$
\end{center}
\includegraphics{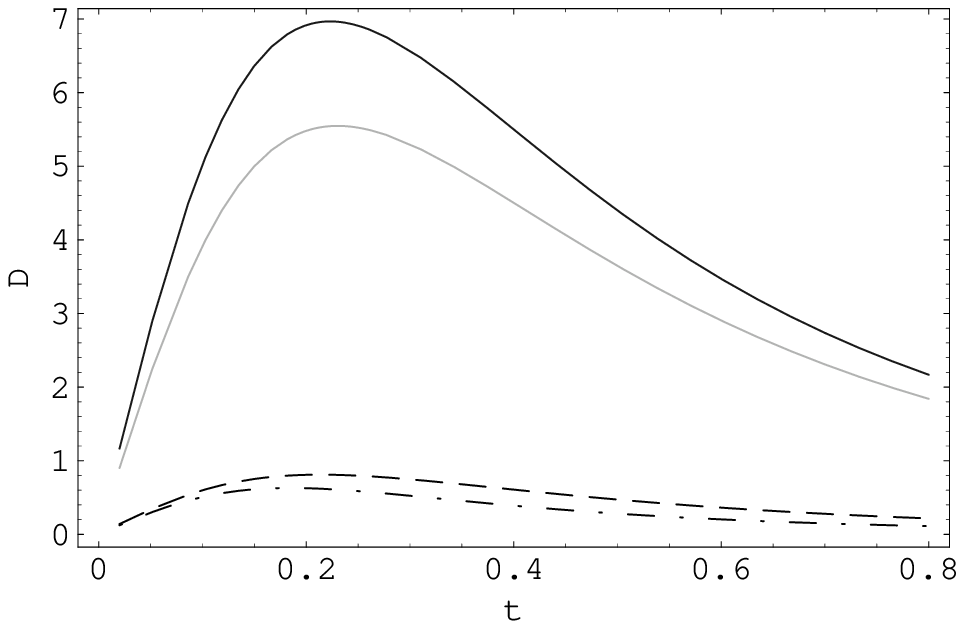}
\vspace{-2mm}
\begin{center}
$\bar b\to\Upsilon$
\end{center}
\includegraphics{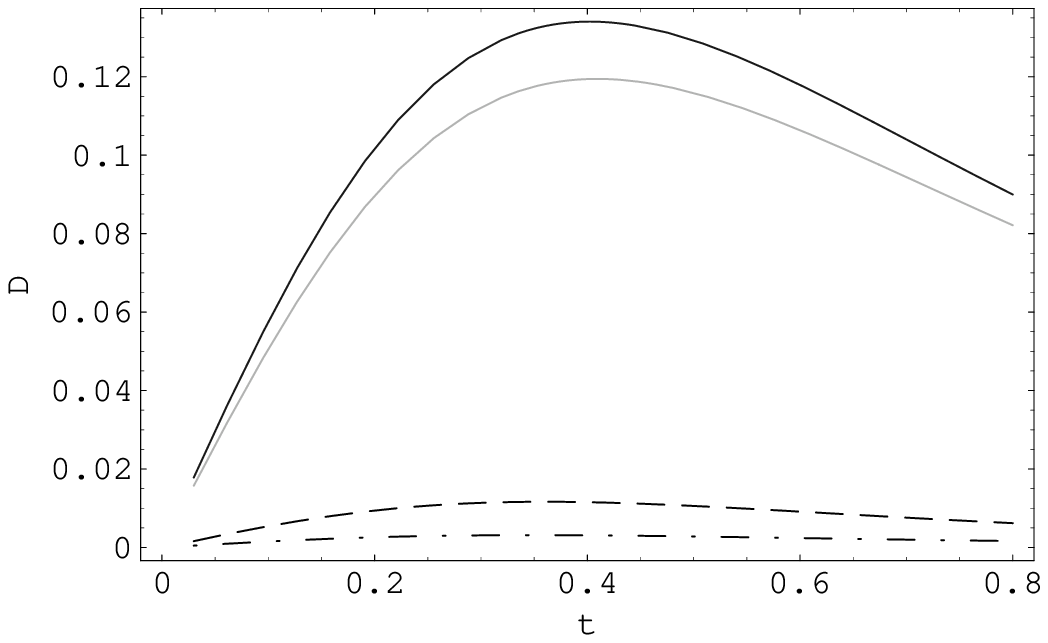}
\vspace{-4mm}
\caption{The contributions to the transverse distributions
$D(\bar c\to J/\Psi)(t)$, $D(\bar b\to B_c^\ast)(t)$,
$D(\bar b\to \Upsilon)(t)$ relative to the heavy quark fragmentation axis.
The thick solid line shows the total fragmentation function, the
dotdashed line shows relativistic correction (37), the dashed line
shows the correction proportional to binding energy $W$ (36). 
The thin solid
line corresponds to the distributions without corrections.
All functions have been multiplied by a factor $10^{4}$.}
\end{figure}

\begin{figure}[htbp]
\vspace{-6mm}
\begin{center}
$\bar c\to \eta_c$
\end{center}
\centering
\includegraphics{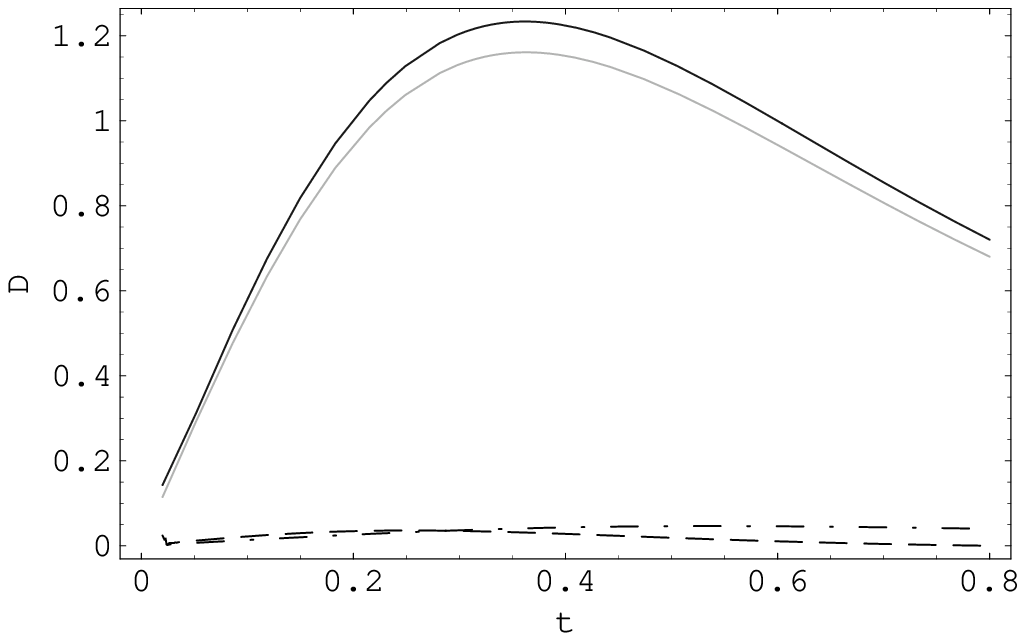}
\vspace{-2mm}
\begin{center}
$\bar b\to B_c$
\end{center}
\includegraphics{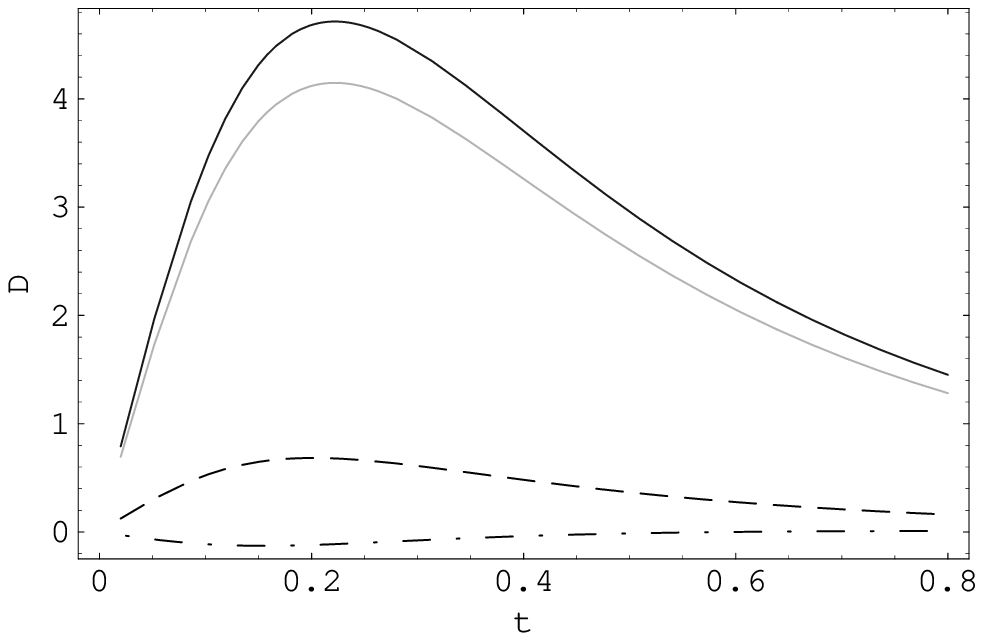}
\vspace{-2mm}
\begin{center}
$\bar b\to\eta_b$
\end{center}
\includegraphics{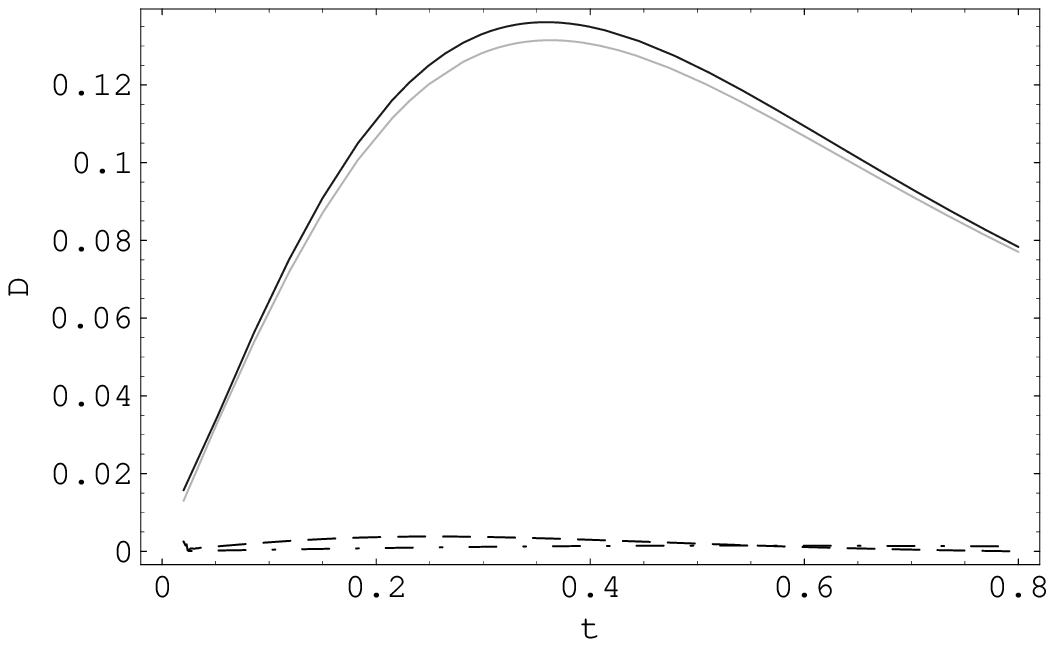}
\vspace{-4mm}
\caption{The contributions to the transverse distributions
$D(\bar c\to \eta_c)(t)$, $D(\bar b\to B_c)(t)$, $D(\bar b\to
\eta_b)(t)$ relative to the heavy quark fragmentation axis. The thick
solid line shows the total fragmentation function, the dotdashed
line shows relativistic correction (53), the dashed line shows the
correction proportional to binding energy $W$ (52). The thin solid
line corresponds to the distributions without corrections.
All functions have been multiplied by a factor $10^{4}$.}
\end{figure}

\begin{figure}[htbp]
\vspace{-6mm}
\begin{center}
$\bar c\to J/\Psi,\eta_c$
\end{center}
\centering
\includegraphics{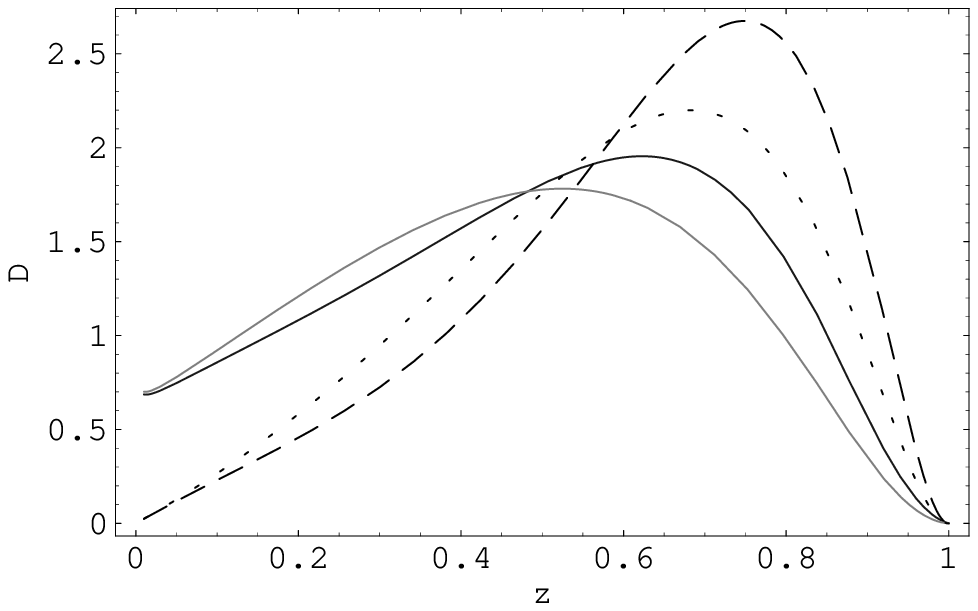}
\vspace{-2mm}
\begin{center}
$\bar b\to B_c^\ast,B_c$
\end{center}
\includegraphics{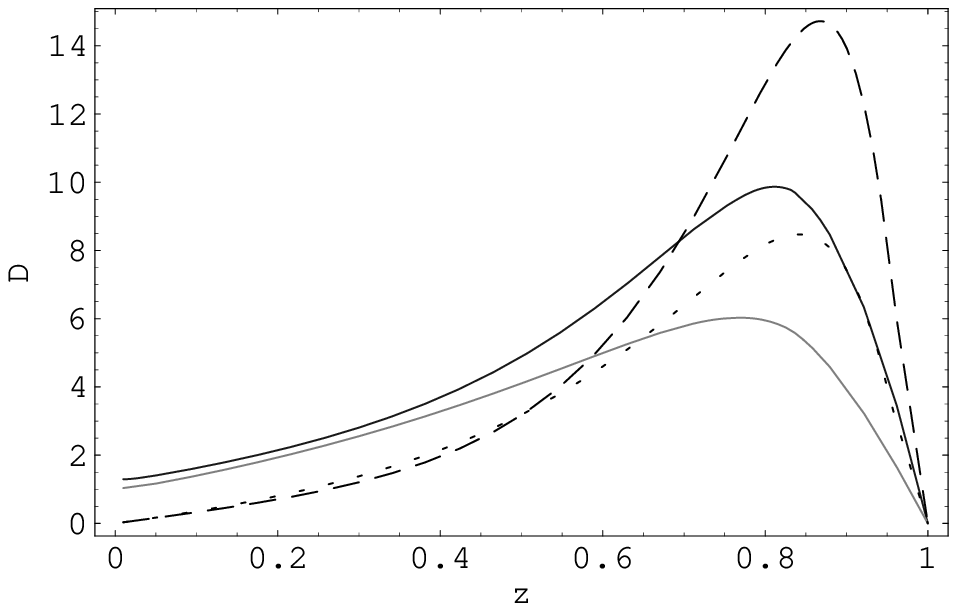}
\vspace{-2mm}
\begin{center}
$\bar b\to\Upsilon,\eta_b$
\end{center}
\includegraphics{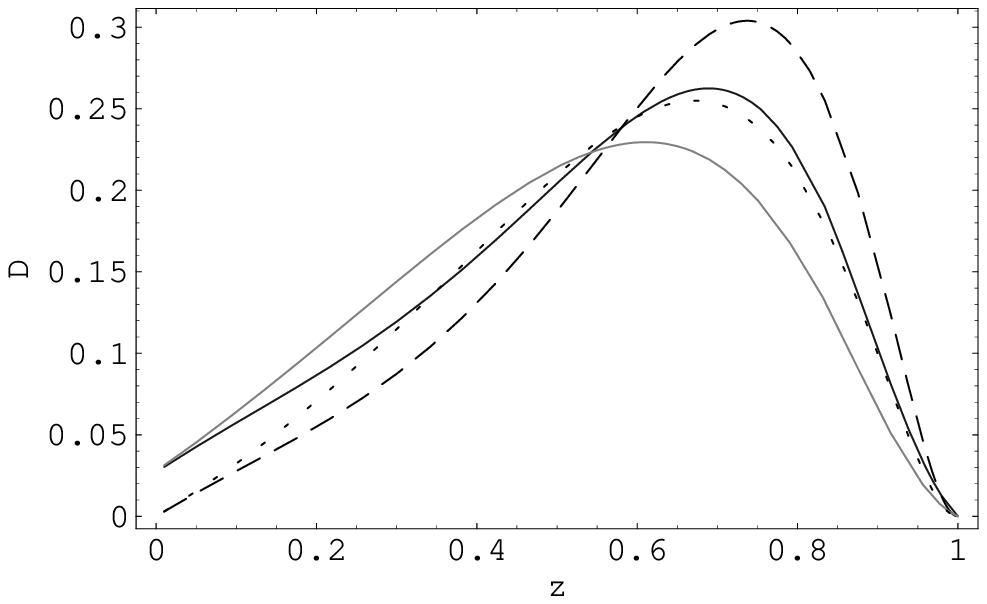}
\vspace{-4mm}
\caption{The total fragmentation functions for the production of
vector and pseudoscalar heavy mesons as a function of $z$ for $\mu=m_1+2m_2$
(dashed lines) and $\mu=M_Z/2$ (solid lines). The thick and thin solid lines
correspond to the vector and pseudoscalar mesons respectively. 
All functions have been multiplied by a factor $10^{4}$.}
\end{figure}

\section{Discussion and Conclusions}

\label{sec:concl}

As mentioned above, the problem of heavy hadron production in $e^+e^-$
and $p\bar p$ collisions became very urgent in last years. The
experimental investigations carried out in this field allowed to measure
the $b$ quark fragmentation function in $Z^0$ decays \cite{Z00}.
The study of the fragmentation processes is important as a tool
to reveal the features of nonperturbative quantum chromodynamics.
There appear experimental data indicating essential differences between
the theoretical predictions and experiment \cite{BFY,RunII,QWG,BC,BLL}.
In the present study we investigated the role of relativistic and bound state
corrections in the heavy quark $b$, $c$ fragmentation processes.
The amplitude of heavy quark fragmentation is obtained in a new form
(13) which accounts all possible relativistic factors for the calculation of 
relativistic corrections to the fragmentation functions. Let us summarize 
several 
peculiarities related to the calculation performed above.

1. We obtain the heavy quark fragmentation functions for both heavy
quarks $b$ and $c$ which fragment to pseudoscalar and vector heavy mesons
starting with the meson production amplitude (1).

2. All possible sources of relativistic corrections including the
transformation
factors for the two quark bound state wave function have been taken into 
account.

3. We investigated the role of relativistic effects in the fragmentation
probabilities over two variables: the longitudinal momentum $z$ and
transverse momentum $p_T$ of the heavy meson.

Analyzing the obtained analytical expressions for the fragmentation
functions both in longitudinal and transverse momentum we can
point out that the calculated corrections for all vector and
pseudoscalar mesons are not exceeding 20 $\%$ the leading order contribution.
The numerical value of the
binding energy correction is dependent on the initial choice of
the heavy quark masses because $W=(M-m_1-m_2)$. So, for example,
the binding energy corrections are extremely small for the
charmonium production in our model where $W_{J/\Psi}=-0.003$ Gev.
The relativistic correction is essentially more important in this
case. Earlier the binding energy and relativistic corrections
were studied in Ref.\cite{Bashir} for the fragmentation
functions of a charm quark to decay into $\eta_c$ and $J/\Psi$.
The comparison of our results with the calculation in
Ref.\cite{Bashir} shows that in the case of the reaction $\bar
c\to J/\Psi$ the binding corrections are numerically close, but relativistic
corrections connected with the expression (60) are essentially
different both in the sign and numerical value. Our relativistic
correction coincides in the sign with the leading order
contribution and is numerically three times smaller than in
Ref.\cite{Bashir}. Moreover, our numerical estimations of analytical relations 
are based on a different numerical value for the expression (60). 
In the paper \cite{Bashir} the parameter $\left\langle{\bf p}^2\right\rangle$
was fixed (using the mass of $c$-quark $m_c=1.43~GeV$) by means of a condition
analogous to our equation (60) without the addendum
proportional to the parameter $B$ entering in the confinement part of
the potential. The results obtained in the present study evidently
show that the relativistic plus bound state corrections lead to the systematic 
increase of the fragmentation probabilities for the pseudoscalar and vector 
mesons. Our total fragmentation functions for the
decays $\bar c\to (c\bar c)$, $\bar b\to (\bar b c)$, $\bar b\to (b\bar b)$ 
retain the initial shape of the leading
order contribution. In the production of vector mesons both
corrections proportional to $W$ and $\left\langle{\bf p}^2\right\rangle$ have 
the same
sign giving us a more essential modification of the leading order
contribution in the comparison with the pseudoscalar meson
production.

The fragmentation functions (19), (27), (42) depend not only on
$z$ but also on the factorization scale $\mu$.
They should be considered at a scale $\mu$ of the order of the heavy
quark masses. The evolution of the fragmentation functions to the
scale $\mu=M_Z/2$ is determined by the DGLAP equation
\cite{DGLAP}:
\begin{equation}
\mu^2\frac{\partial}{\partial\mu^2}D_{Q\to
H}(z,\mu^2)=\int_z^1\frac{dy}{y}P_{Q\to
Q}\left(\frac{z}{y},\mu\right)D_{Q\to H}(y,\mu^2),
\end{equation}
where $P_{Q\to Q}(x)$ is the quark splitting function \cite{LP}.
The modification of the z-shape of the fragmentation functions is
shown in Fig.6. The average values of the momentum fraction for
the production of different heavy mesons at the scale $\mu=M_Z/2$
are the following: $<z>=0.50$ $(J/\Psi)$, $<z>=0.46$ $(\eta_c)$,
$<z>=0.63$ $(B^\ast_c)$, $<z>=0.59$ $(B_c)$, $<z>=0.56$
$(\Upsilon)$, $<z>=0.52$ $(\eta_b)$.

In the case of $(\bar c c)$ or $(\bar b b)$ mesons Eqs.(29) and
(46) acquire a more simple form:
\begin{equation}
\Omega_V=\int_0^1D_{\bar Q\to V(\bar 
QQ)}(z)dz=\frac{32\alpha_s^2|\Psi(0)|^2}{27m_2^3}
\Bigl[\frac{1189}{30}-57\ln 2+
\end{equation}
\begin{displaymath}
+\frac{W}{m_2}\left(134\ln 2-\frac{78149}{840}\right)+
\frac{\left\langle{\bf p}^2\right\rangle}{m_2^2}\left(\frac{1078}{9}\ln 
2-\frac{78416}{945}\right)\Bigr].
\end{displaymath}
\begin{equation}
\Omega_P=\int_0^1D_{\bar Q\to P(\bar 
QQ)}(z)dz=\frac{8\alpha_s^2|\Psi(0)|^2}{81m_2^3}
\Bigl[\frac{1546}{5}-444\ln 2+
\end{equation}
\begin{displaymath}
+\frac{W}{m_2}\left(-104\ln 2+\frac{15581}{240}\right)+
\frac{\left\langle{\bf p}^2\right\rangle}{m_2^2}\left(\frac{16}{9}\ln 
2-\frac{139}{48}\right)\Bigr].
\end{displaymath}
Numerical values of the total fragmentation probabilities are presented in
Table I. The evolution conserves the integral probabilities $\Omega_V$ and
$\Omega_P$ of the fragmentation. Using expressions (29) and (46) we obtain
the significant experimental ratio
\begin{equation}
\eta(r)=\frac{\Omega_V}{\Omega_V+\Omega_P},
\end{equation}
which predicts the relative number of vector and pseudoscalar
mesons. For the $(\bar c c)$, $(\bar b c)$ and $(\bar bb)$ mesons
the ratio (64) gives the following numbers: 0.46, 0.58, 0.52. The obtained
results for the relativistic and bound state corrections to the different
heavy quark fragmentation functions are shown in Figs.2-5. Relative order 
contributions of relativistic
and binding corrections are the biggest for the $(\bar b c)$
mesons because of the little growth the parameter (60) and bound state
energy $W$ as compared to $(c\bar c)$ states. The decrease of these corrections
in the bottomonium is explained by the increase of heavy quark mass ($m_c\to 
m_b$). 
All considered effects in the production of $(\bar b c)$,
$(\bar b b)$ and $(\bar c c)$ mesons are computed to the leading
order in $\alpha_s$ in the color singlet model. For S-states like
$J/\Psi$, $\eta_c$, $B_c$, $B_c^\ast$, $\eta_b$ and $\Upsilon$
the color-octet terms are suppressed relative to the color
singlet terms by a factor of the second order over the relative
velocity $v_Q$ \cite{BFY}.
Our results should be useful for the comparison
with more accurate $(\bar c c)$, $(\bar b b)$ and $(\bar b c)$
meson production measurements in $Z^0$ decays or in $p\bar p$
collisions at the Tevatron.

\acknowledgments
I am grateful to D.Ebert, R.N.Faustov, V.O.Galkin for a careful reading of the 
manuscript,
useful remarks and suggestions, and to I.B.Khriplovich, V.V.Kiselev, 
A.K.Likhoded, V.A.Saleev
for useful discussions of different questions regarding this study. 
The author thanks the colleagues from the
Institute of Physics of the Humboldt University in Berlin for warm
hospitality.
The work is performed under the financial support of the
{\it Deutsche Forschungsgemeinschaft} under contract Eb 139/2-3.

\end{document}